\documentclass[aps,twocolumn,prx,superscriptaddress]{revtex4-2}
\usepackage{blindtext}
\usepackage{physics,siunitx}
\usepackage{standalone}
\usepackage{floatrow}
\usepackage{amsmath,amsfonts,amssymb}
\usepackage{graphicx}
\usepackage{multirow}

\usepackage{hyperref}
\usepackage{soul}
\usepackage{xcolor}

\begin{document}
	\title{Voltage Activated Parametric Entangling Gates on Gatemons
 }
	
	\author{Yinqi Chen}
    \affiliation{Department of Physics  and Wisconsin Quantum Institute, University of Wisconsin-Madison, Madison, Wisconsin 53706, USA}
	
	\author{Konstantin N. Nesterov}
    \affiliation{Department of Physics  and Wisconsin Quantum Institute, University of Wisconsin-Madison, Madison, Wisconsin 53706, USA}
    \affiliation{Bleximo Corp., Berkeley, California 94710, USA}

\author{Hugh Churchill}
    \affiliation{Department of Physics, University of Arkansas, Fayetteville, Arkansas 72701, USA}

    \author{Javad Shabani}
    \affiliation{Center for Quantum Information Physics, Department of Physics, New York University, NY 10003, USA}

\author{Vladimir E. Manucharyan}
    \affiliation{Department of Physics, Joint Quantum Institute, and Quantum Materials Center, University of Maryland, College Park, MD 20742, USA}
    \affiliation{EPFL, CH-1015 Lausanne, Switzerland}

\author{Maxim G. Vavilov}
    \affiliation{Department of Physics and Wisconsin Quantum Institute, University of Wisconsin-Madison, Madison, Wisconsin 53706, USA}

    \date{\today}
    
\begin{abstract}
We describe the generation of entangling gates on superconductor-semiconductor hybrid qubits by ac voltage modulation of the Josephson energy. Our numerical simulations demonstrate that the unitary error can be below $10^{-5}$ in a variety of 75-ns-long two-qubit gates (CZ, $i$SWAP, and $\sqrt{i\mathrm{SWAP}}$) implemented using parametric resonance.  We analyze the conditional $ZZ$ phase and demonstrate that the CZ gate needs no further phase correction steps, while the $ZZ$ phase error in SWAP-type gates can be compensated by choosing pulse parameters. With decoherence considered, we estimate that qubit relaxation time needs to exceed $\SI{70}{\micro\second}$ to achieve the 99.9\% fidelity threshold.
\end{abstract}

\maketitle

\section{Introduction}
Two-qubit gates present a challenge in the development of scalable superconducting quantum processors. In the simplest hardware architectures based on fixed-frequency qubits and always-on couplings, two-qubit gates can be realized by microwave control~\cite{Paraoanu_2006, Chow_2013, Krinner_2020, Mitchell_2021} via, for example, the cross-resonance effect~\cite{Chow_2011, Sheldon_2016, Kandala_2021}. Nevertheless, gates with fixed-frequency transmon qubits~\cite{Koch2007} and fixed couplings lead to stringent frequency allocation conditions~\cite{Morvan_2021}, which  complicate processor design and reduce the yield of multi-qubit chips in microfabrication. An alternative way is to make qubit frequencies tunable so that two-qubit states can be moved into and out of resonance conditions to induce transitions between qubit states or to enable conditional phase accumulations~\cite{DiCarlo_2009, Yamamoto_2010, Chen_2014, McKay_2015, Barends_2019}. 

While the tunability adds more freedom for gate and processor design, it may reduce the processor's performance due to crowded spectrum of multiqubit systems~\cite{Schutjens_2013}. During the frequency tuning process, one can cross unintended resonances, making it hard to control specific pairs of qubits without inducing spectator errors in other qubits.
One approach to avoid crossing undesirable resonances is to induce a parametric resonance on a desired pair of qubits. The latter can be achieved by modulating qubit frequencies~\cite{Beaudoin_2011, Strand_2013, Naik_2017, Didier2018, Caldwell2018, Reagor_2018, Hong2020, Sete2021} or coupling elements~\cite{Bertet_2006, Niskanen_2006, Niskanen2007, McKay2016, Royer_2017, Ganzhorn2020, Valery_2022}.
The parametric resonance response is  determined by the modulation amplitudes and frequencies, thus one can target a specific pair  in an array of qubits with similar frequencies. For flux-tunable transmon qubits, frequency modulation has been realized by varying the external flux bias through the SQUID loop, which changes the effective Josephson energy~\cite{Caldwell2018, Didier2018, Valery_2022}. 

Another way to control the qubit characteristics is to use voltage tunable qubits, so-called "gatemons," where the metallic gate voltage controls the charge carrier density in the Josephson junctions~\cite{Larsen2015,de_Lange_2015, Casparis_2018, Hertel2022}. 
The gate voltage control eliminates flux noise sensitivity~\cite{Casparis_2018} and suppresses charge dispersion~\cite{Krinhoj_2020}. Despite these advantages, gatemons are known for relatively low coherence time, which limits the achievable fidelity of gate operations. For the successful realization of two-qubit gates, a certain threshold on coherence has to be met.

In this paper, we analyze the possibility of implementing entangling gates on gatemon qubits by voltage modulation of the Josephson energy and estimate the requirements for gatemon coherence. We look at three types of entangling gates that belong to different classes of local equivalence: \emph{(i)} the CZ gate, \emph{(ii)} the $i$SWAP gate, and \emph{(iii)} the $\sqrt{i\mathrm{SWAP}}$ gate. We computationally show that all three gates can be implemented with fidelity $>99.99\%$ for unitary evolution. Once  decoherence is taken into consideration, the fidelity greater than $99.9\%$ requires that the relaxation time $T_1$ exceeds $\SI{70}{\micro\second}$. 

We analyze the error budget and identify several primary coherent-error mechanisms for the two-qubit gates. In particular, we find that the unwanted conditional $ZZ$ phase is the most harmful error for SWAP-type gates and propose mitigation of this error using qubit tunability. Because of its simplicity, this mitigation approach is beneficial in comparison to resorting to advanced pulse shape designs~\cite{Sudaresan_2020} or to tunable couplers~\cite{Mundada_2019, Collodo_2020}, which add complexity to the control scheme. For the $i$SWAP gate, we demonstrate that the accumulated $ZZ$ phase can be reduced to zero by tuning the static Josephson energy for an experimentally accessible range of the coupling constant.
For $\sqrt{i\mathrm{SWAP}}$, the parameter range of suppressed phase error is wider and can be further expanded  by using a two-tone modulation pulse with close frequencies.

The paper is organized as follows. In Sec.~\ref{Sec:basic}, we introduce the Hamiltonian of interacting gatemon qubits that we use to model entangling gates. In Sec.~\ref{Sec:cz} we present the simulation of the CZ gate, showing that the CZ gate can be optimized to high fidelity $>99.99\%$ for a broad range of coupling constant $J_C$. In Sec.~\ref{Sec:iswap}, we present the simulation of the $i$SWAP gate, analyze the detrimental effect of $ZZ$ phase on the gate fidelity, and discuss how this phase can be compensated by choosing appropriate static Josephson energy. In Sec.~\ref{Sec:sqrtiswap}, we present the simulation of the  $\sqrt{i\mathrm{SWAP}}$ gate and similarly discuss how the $ZZ$ phase can be compensated. We do this for both one-tone and two-tone pulse schemes.  In Sec.~\ref{Sec:decoherence}, we evaluate the effect of decoherence on gate fidelity and discuss requirements on the relaxation time $T_1$ of future-generation gatemon qubits to enable high-fidelity gates.

\section{Basic Model}\label{Sec:basic}

\subsection{Qubit Hamiltonian}\label{Sec:Gatemon}

The idea of implementing two-qubit gates by directly modulating qubit spectra has been proposed and realized on transmon qubits~\cite{Caldwell2018, Reagor_2018, Didier2018, Hong2020, Sete2021}.  With the proper choice of frequencies, modulation pulses can selectively induce Rabi oscillations between desired states to generate gates. For transmons, the qubit-frequency modulation is achieved by tuning the external magnetic field. Here we  apply the idea of parametric control onto gatemon qubits using gate voltage as the control instead of the external magnetic field.

\begin{figure}
    \centering
    \includegraphics[width=\textwidth]{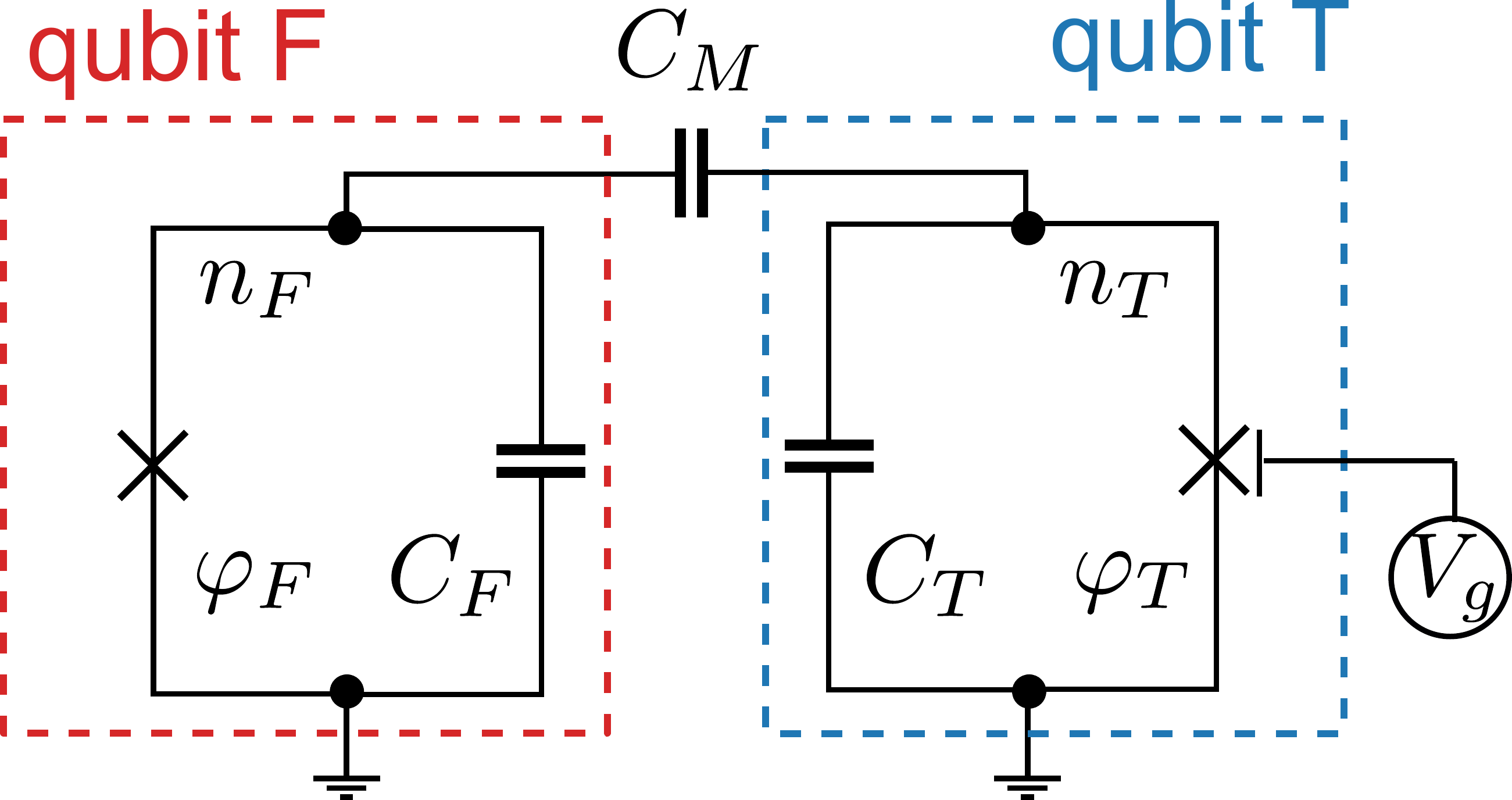}
    \caption{Circuit diagram of two capacitively coupled gatemon qubits. The Josephson junction of the tunable qubit (T) is gated by a voltage $V_g$, which can be used to control the Josephson energy $E_{J,T}$ of the tunable qubit.}
    \label{fig:circuit}
\end{figure}

In a gatemon qubit, the Josephson junction (JJ) is implemented with a superconductor-normal-superconductor (SNS) structure, where the normal section is a semiconductor. The semiconductor allows Josephson energy $E_J$ to be tuned by the gate voltage $V_g$~\cite{Larsen2015,de_Lange_2015,Aguado_2020,Hertel2022}. The quantum system then consists of a fixed qubit (F) and a gate-tunable qubit (T) with capacitive coupling, as shown in Fig.~\ref{fig:circuit}. The Hamiltonian of this circuit is
\begin{equation}\label{eq:2qbit}
    \hat{H} = \hat{H}_F + \hat{H}_T + J_C\hat{n}_F\hat{n}_T\,,
\end{equation}
where $\hat{H}_i$ ($i=F, T$) are single qubit Hamiltonians
\begin{equation}\label{eq:1qbit}
    \hat{H}_i = 4E_{C,i}\hat{n}_i^2-E_{J,i}\cos(\hat{\varphi}_i)\,,
\end{equation}
and $\hat{n}_i,\hat{\varphi}_i$ are the charge and phase operators. The charge coupling constant $J_C$ is determined by the capacitances of circuit components. [It is also common to write the coupling term as the exchange coupling $-g(a^{\dagger}-a)(b^{\dagger}-b)$. The two constants are related by the formula $g=4J_C(E_{J,F}E_{J,T}/4E_{C,F}E_{C,T})^{1/4}$].  The parameters $E_{C,i}$ and $E_{J,i}$ are respectively the charging  and Josephson energies, which  obey the standard transmon condition $E_{J,i}\gg E_{C,i}$~\cite{Koch2007}. For the gatemon qubit considered here, $E_{J,T}$ is not controlled by flux bias, but instead by the gate voltage $V_g$:
\begin{equation}
    E_{J,T} = E_{J,T}(V_g).
\end{equation}
For an actual device, the dependence of $E_{J}$ on $V_g$ does not have a simple relation. However, we will assume a first-order approximation in the current theoretical study:
\begin{equation}\label{eq:EJgate}
    E_{J,T}(V_0+\delta V_g) = E_{J,T}(V_0)+E_{J,T}'(V_g)\Big|_{V_g=V_0}\delta V_g\,.
\end{equation}
A sinusoidal pulse of voltage will induce sinusoidal oscillation of $E_{J,T}$
\begin{equation}\label{eq:pulse}
    E_{J,T}(t)=\bar{E}_{J,T}+f(t)\delta E_{J,T}\cos(\omega_pt)\,,
\end{equation}
where $\omega_p$ is the modulation frequency, $\delta E_{J,T}$ is the amplitude of the modulation, and $\bar{E}_{J,T}$ is the \textit{static} Josephson energy. We use a Gaussian flat-top 
envelope $f(t)$ defined as
\begin{equation}
    f(t)=\begin{cases}
    \exp(-\dfrac{2(t-t_{\rm left})^2}{t_{\rm rise}^2}) - e^{-2} & 0< t < t_{\rm left}\\
    1 - e^{-2}  & t_{\rm left} < t < t_{\rm right} \\
    \exp(-\dfrac{2(t-t_{\rm right})^2}{t_{\rm rise}^2}) - e^{-2}  & t_{\rm right}<t < t_{\rm gate} \,,
    \end{cases}
\end{equation}
where $t_{\rm left}=t_{\rm rise}$, $t_{\rm right}=t_{\rm gate}-t_{\rm rise}$. Here $t_{\rm rise}$ is the ramping time for the pulse to ramp up and down, and $t_{\rm gate}$ is the gate time. 

We denote the fixed-qubit frequency by $\omega_F$, and write the time-dependent tunable-qubit frequency $\omega_T$ in the presence of the voltage modulation \eqref{eq:pulse} as the Fourier series:
\begin{equation}\label{eq:harmonics}
\omega_T(t) = \bar{\omega}_T + \sum_{m=1}^{\infty} \delta\omega_m\cos(m\omega_p t)\,.
\end{equation}
We also write $\eta_F>0$ for the absolute value of the fixed-qubit anharmonicity and use $\bar{\eta}_T>0$ for the time-averaged anharmonicity of the tunable qubit.

In numerical simulations, we use the following qubit parameters: $E_{C,F}/h=E_{C,T}/h=\SI{0.2}{\giga\hertz}$, $E_{J,F}/E_{C,F}=100, \bar{E}_{J,T}/E_{C,T}=78$. We simulate all three types of entangling gates for gate time fixed at $\SI{75}{\nano\second}$. The corresponding qubit frequencies and anharmonicities without modulation are $\omega_F/(2\pi) = \SI{5.449}{\giga\hertz}$, $\eta_F/(2\pi)=\SI{0.219}{\giga\hertz}$, $\bar{\omega}_T/(2\pi)=\SI{4.787}{\giga\hertz}$, $\bar{\eta}_T/(2\pi)=\SI{0.222}{\giga\hertz}$. For these system parameters, the structure of the static single-qubit spectra  can be inferred from Table~\ref{tab:single_e}. The coupling $J_C$ ranges from $\SI{10}{\mega\hertz}$ to $\SI{30}{\mega\hertz}$. 

As preparation for later discussion, we mention that the coupling between two qubits will introduce $ZZ$ interaction even without any driving. 
The static $ZZ$ phase accumulation rate due to this interaction is $(E_{\overline{00}}+E_{\overline{11}}-E_{\overline{01}}-E_{\overline{10}})/h$, where $E_{\overline{ij}}$ are dressed eigenenergies. For example, when $J_C=\SI{10}{\mega\hertz}$ and $\bar{E}_{J,T}/E_{C,T}=78$, the $ZZ$ phase accumulation rate is $\SI{1.41}{\mega\hertz}$. 

\begin{table}
    \centering
\begin{tabular}{c|c|c}
    \hline\hline
    Transition Process  & Fixed Qubit & Tunable Qubit\\
    \hline
    $0\rightarrow 1$ & 5.449 & 4.787\\
    $1\rightarrow 2$ & 5.230 & 4.565\\
    $2\rightarrow 3$ & 4.995 & 4.323\\
    \hline\hline
\end{tabular}
\caption{Static transition frequencies of single qubits. Units are in GHz.}
\label{tab:single_e}
\end{table}

\subsection{Parametric-Resonance Conditions}\label{Sec:Gates}

While we use the full circuit Hamiltonian \eqref{eq:2qbit} for simulations, in this section, we identify relevant terms in the interaction-picture Hamiltonian, which dominate system dynamics for specific gates~\cite{Didier2018, Caldwell2018}. 
The  choice of $\omega_p$ determines the resonance condition. For iSWAP processes,  when $\omega_p\approx\Delta$, the relevant term is
\begin{subequations}\label{eq:int}
\begin{equation}
\label{eq:intSWAP}
    \hat{H}_{i\mathrm{SWAP}} = g_{10,01}^{(1)}
    e^{-i(\omega_p-\Delta) t}\ket{10}\bra{01}+\mathrm{h.c.}\,.
\end{equation}
For CZ processes, the relevant term is
\begin{equation}\label{eq:H_CZ1}
     \hat{H}_{\rm CZ}^{(1)} = g_{20,11}^{(1)}e^{i(\omega_p-\Delta-\eta_F)t}\ket{20}\bra{11}+\mathrm{h.c.}\,
\end{equation}
when $\omega_p\approx \Delta+\eta_F$ or
\begin{equation}\label{eq:H_CZ2}
    \hat{H}_{\rm CZ}^{(2)} = g_{11,02}^{(1)}e^{i(\omega_p-\Delta+\bar{\eta}_T)t}\ket{11}\bra{02}+\mathrm{h.c.}\,
\end{equation}
\end{subequations}
when $\omega_p\approx\Delta-\bar{\eta}_T$. Here $g_{ij,kl}^{(1)}$ is the renormalized interaction matrix element,
$\Delta=\bar{\omega}_T-\omega_F$ is the detuning between the time average of tunable  and fixed qubit frequencies, and  $\ket{ij} = \ket{i}_F\otimes \ket{j}_T$ is the two-qubit product state corresponding to qubit $F$ being in state $\ket{i}_F$ and $T$ being in state  $\ket{j}_T$. A more detailed discussion of the Hamiltonian terms in Eqs. \eqref{eq:int} can be found in Appendix~\ref{App:int}.

We point out here that for the experiment with transmon qubit,  parametric modulation is usually operated at the sweet spot, therefore only even-$m$ harmonics are present in the corresponding analog of the Fourier series~\eqref{eq:harmonics}~\cite{Caldwell2018, Sete2021}. For demonstration in this paper, on the other hand, we choose to operate at a point where the frequency linearly depends on voltage, and thus all harmonics are present in Eq.~\eqref{eq:harmonics}.

\subsection{Two-qubit Gates}

The two-qubit entangling gates can be classified by their local equivalence, i.e., whether they can be transformed into each other solely using single-qubit operations. Such classes of equivalence relations can be characterized by two local invariants~\cite{Zhang_2003}. We analyze the two-qubit gates in the following form~\cite{Nesterov_2021}:
\begin{equation}
\label{eq:Uid}
\hat{U}_{\rm ideal}(\theta,\zeta)=\begin{pmatrix}
e^{-i\zeta/2} & 0 & 0 & 0 \\
0 & \cos\frac\theta 2 & -i\sin\frac\theta 2 & 0 \\
0 & -i\sin\frac\theta 2 & \cos\frac\theta 2 & 0 \\
0 & 0 & 0 & e^{-i\zeta/2}
\end{pmatrix}\,.
\end{equation}
In particular, we look at the three cases when
\begin{subequations}
\begin{equation}\label{CZ_angles}
    \zeta=\pi,\quad \theta=0\,,
\end{equation}
which is locally equivalent to the CZ gate,
\begin{equation}\label{ISWAP_angles}
    \zeta=0,\quad \theta=\pi\,,
\end{equation}
which is the $i$SWAP gate, and
\begin{equation}\label{sqISWAP_angles}
    \zeta=0,\quad \theta=\pi/2\,,
\end{equation}
\end{subequations}
which is the $\sqrt{i\mathrm{SWAP}}$ gate. By Eqs.~\eqref{eq:int}, the resonance condition for Eqs.~\eqref{ISWAP_angles} and \eqref{sqISWAP_angles}  is $\omega_p=\Delta$, while for Eq.~\eqref{CZ_angles}, the condition is either $\omega_p=\Delta-\bar{\eta}_T$ or $\omega_p=\Delta+\eta_F$. 

We assume that we can apply virtual single-qubit $Z$ rotations before and after the pulses so that the evolution operator in the computational subspace can be locally transformed to the form given by Eq.~\eqref{eq:Uid}~\cite{Chow2017}. Therefore, the evolution operator for gate-fidelity calculations has the form:
\begin{equation}
    \hat{U}=\hat{U}_{\rm post}\hat{U}_{\rm pulse}\hat{U}_{\rm pre}\,,
\end{equation}
where $\hat{U}_{a}=\exp(i\hat{Z}_1\vartheta_{a,1})\exp(i\hat{Z}_2\vartheta_{a,2})$ for $a=\mathrm{pre},\mathrm{post}$ are the local $Z$ rotations, and $\hat{U}_{\rm pulse}$ is the evolution induced by the pulse  Eq.~\eqref{eq:EJgate} and projected into the computational subspace. The coherent gate fidelity is then calculated using the final  matrix $\hat{U}$ as~\cite{Pedersen_2007}
\begin{equation}\label{eq:fidelity}
    F= \dfrac{\Tr(\hat{U}^{\dagger}\hat{U})+\abs{\Tr[\hat{U}_{\rm ideal}(\theta,\zeta)^{\dagger}\hat{U}]}^2}{20}\,.
\end{equation}

\section{The CZ Gate}\label{Sec:cz}
The CZ gate can be achieved by satisfying the resonance condition $\omega_p=\Delta+\eta_F$ of $\hat{H}_{\rm CZ}^{(1)}$, see Eq.~\eqref{eq:H_CZ1}, or $\omega_p=\Delta-\bar{\eta}_T$ of $\hat{H}_{\rm CZ}^{(2)}$, see Eq.~\eqref{eq:H_CZ2}. Here we choose the latter condition for simulation. With this choice of $\omega_p$, a $2\pi$-rotation in the $\ket{11}-\ket{02}$ subspace transforms state $\ket{11}$ into $e^{i\pi}\ket{11}$, thus inducing a CZ gate.

\begin{center}
\begin{figure}
    \centering
    \includegraphics[width=0.8\textwidth]{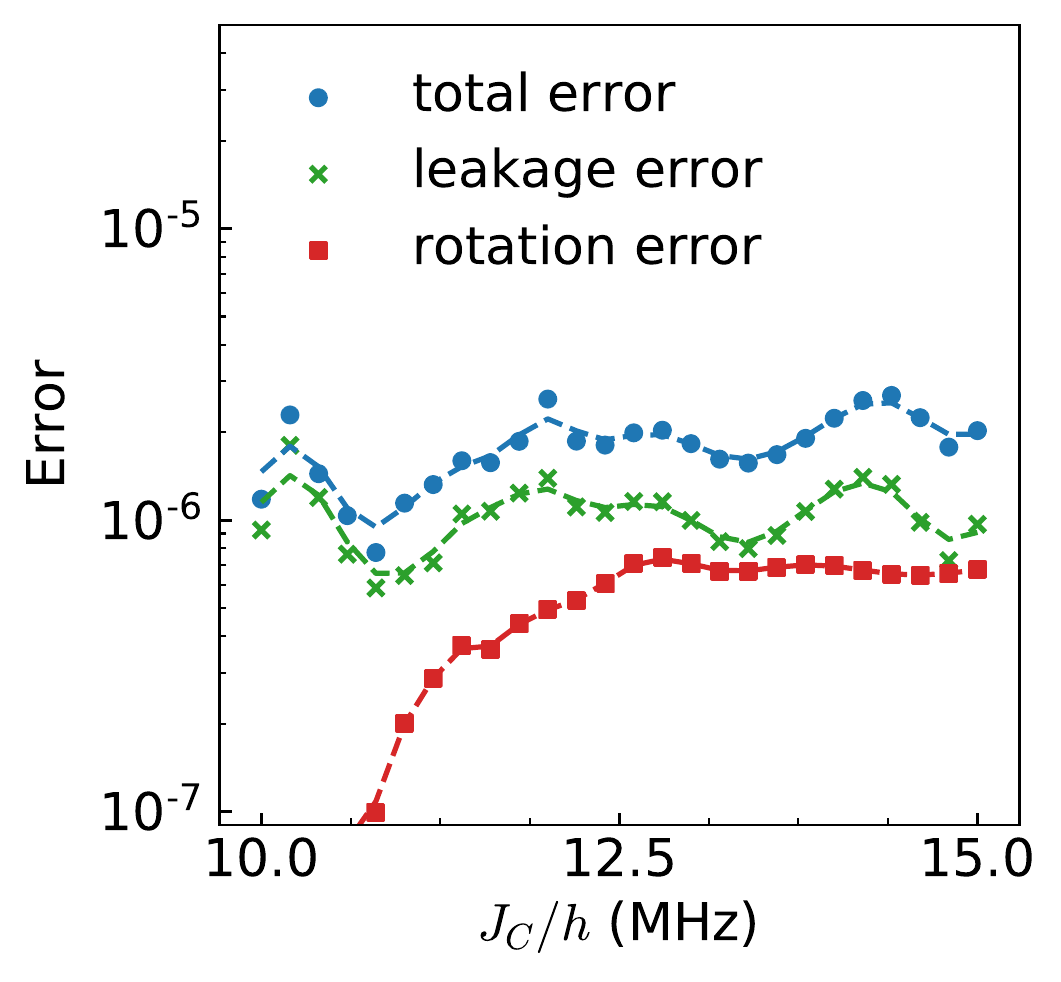}
    \caption{The total coherent error of the CZ gate (blue dots), together with estimates of the leakage error (green crosses) and rotation error (red squares). The dashed lines correspond to smooth fits and are drawn for visual clarity.  The qubit parameters are chosen to be $E_{C,F}/h=E_{C,T}/h=\SI{0.2}{\giga\hertz}$, $E_{J,F}/E_{C,F}=100, E_{J,T}/E_{C,T}=78$.}
    \label{fig:cz_chara}
\end{figure}
\end{center}

The evolution of the system is defined by additional terms of the full Hamiltonian that are presented in Apendix~\ref{App:int}.  Although these terms result in small corrections, we take them into account in numerical simulations.  Here we briefly identify these terms. \emph{(i)} There is small off-resonant phase accumulation for states $\ket{00},\ket{01},\ket{10}$ due to ac Stark effect, see  Appendix~\ref{App:phase}.
\emph{(ii)} The transitions $\ket{00}\leftrightarrow \ket{11}$, $\ket{01}\leftrightarrow \ket{12}$, $\ket{10}\leftrightarrow \ket{21}$ are processes that simultaneously absorb or emit two photons, thus, have largely off-resonant transition frequencies  providing small corrections.
\emph{(iii)} The interaction between two computational states $\ket{01}$ and $\ket{10}$ shifts the state energies with nearly equal magnitudes in opposite directions making small contribution to the $ZZ$ conditional phase $\zeta=\phi_{00}+\phi_{11}-\phi_{01}-\phi_{10}$. Therefore, the overall conditional phase $\zeta$ is mainly determined by the phase accumulation for state $\ket{11}$. This allows us to implement the  CZ gate by a close-to-resonance $2\pi$-rotation between the states $\ket{11}$ and one of the states $\ket{02}$ or $\ket{20}$. We find consistently high-fidelity CZ gate across different couplings $J_C$, as shown in Fig.~\ref{fig:cz_chara} for  $\SI{10}{\mega\hertz}\leq J_C/h\leq \SI{15}{\mega\hertz}$.

We also analyze  the error budget for the CZ gate. The phase error is modeled as an extra phase $\delta\phi$ on the $ZZ$ phase by $\zeta = \pi+\delta\phi$. When no leakage is present, the phase error is approximately $3(\delta\phi)^2/20$, according to equation \eqref{eq:fidelity}. The leakage error is modelled as a leakage in the $\ket{11}$ state $\hat{U}\ket{11}=-\cos\epsilon\ket{11}+i\sin\epsilon\ket{02}$, which gives an error to be approximately $\sin^2(\epsilon)/4$ by \eqref{eq:fidelity}. There could also be unwanted rotation between $\ket{01}$ and $\ket{10}$ states, which we model as $\hat{U}\ket{01}=\cos\gamma\ket{01}+i\sin\gamma\ket{10}$, and similarly for $\ket{10}$ state. The rotation error is approximately $2\sin^2(\gamma)/5$. 

As Fig.~\ref{fig:cz_chara} shows, the leakage error is the dominant contribution to the total error. The rotation error is a few orders of magnitude smaller than the leakage and total error but then increases to around $5\times10^{-7}$ as $J_C$ increases. The phase error contributes the least, mostly around or below $10^{-7}$, and thus is not shown in the plot. This low phase error has a simple interpretation.  The CZ gate is based on a geometric phase induced by the pulse and this geometric phase could be easily tuned by choosing a slightly different trajectory on the 11-02 Bloch sphere to account for additional spurious contributions the conditional phase. For example, the drive can be chosen slightly off-resonance to account for extra phases accumulated by other states.

We note that the fluctuations of the leakage error and the phase error are mainly the results of how the numerical optimization algorithm finds the optimal fidelity rather than some intrinsic mechanism of the gate. The total error of the CZ gate overall is consistently less than $10^{-5}$, which is below the threshold to perform error correction codes that use CZ gates to build stabilizer sequences, e.g., the distance-three code~\cite{Krinner_2022}.

\begin{center}
\begin{figure}
    \centering
    \includegraphics[width=\textwidth]{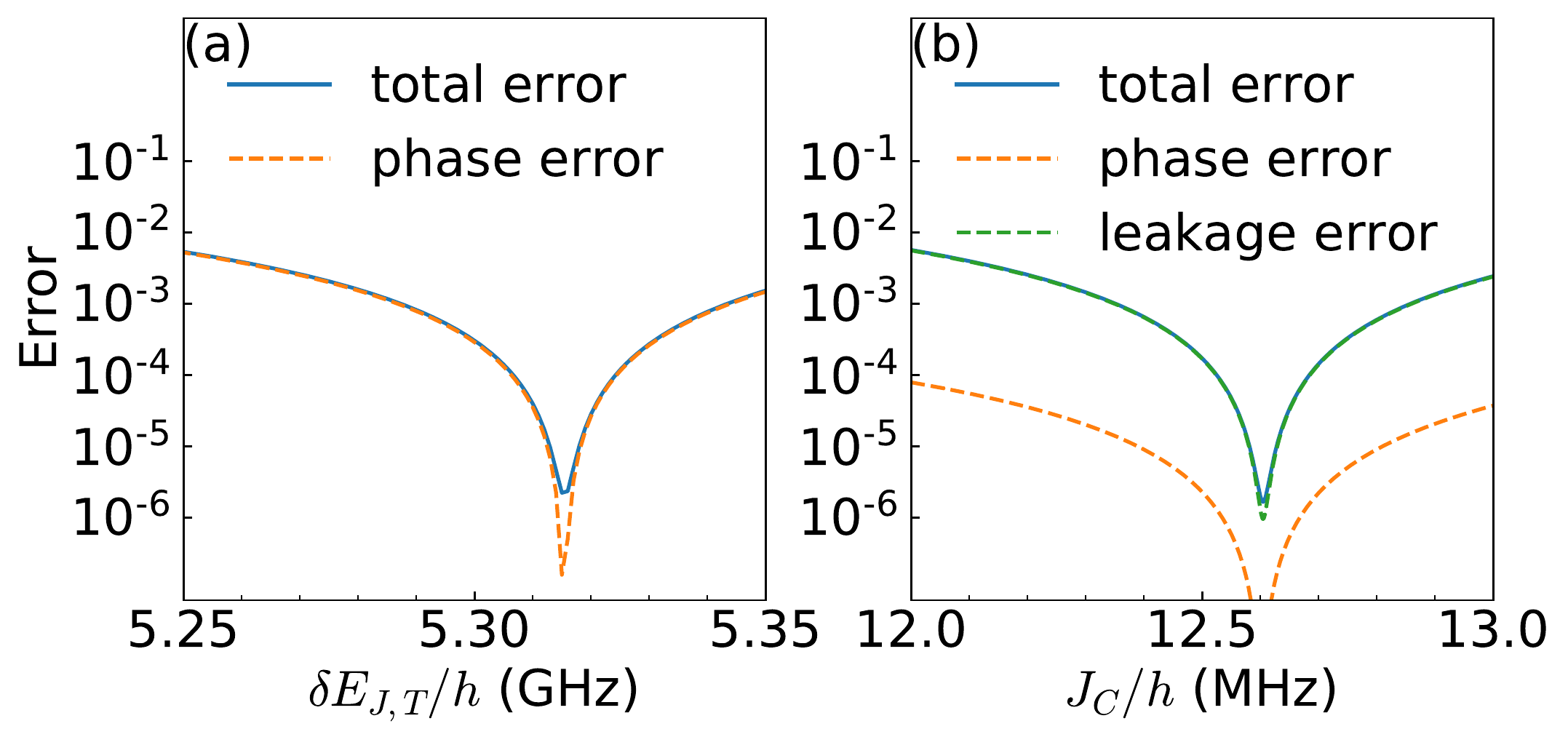}
    \caption{The coherent error of CZ gate (blue solid) under the variation of (a) the drive amplitude $\delta E_{J,T}$ and (b) coupling constant $J_C$, decomposed into the phase error (orange dashed) and leakage error (green dashed). The pulse parameters are optimized at $J_C/h=\SI{12.6}{\mega\hertz}$, as in Fig.~\ref{fig:cz_chara}.}
    \label{fig:cz_fluctuation}
\end{figure}
\end{center}

In Fig.~\ref{fig:cz_fluctuation} we show the sensitivity of the CZ gate fidelity to the variations of the drive amplitude $\delta E_J$ and the coupling constant $J_C$. The pulse parameters used are optimized at $J_C/h=\SI{12.6}{\mega\hertz}$ as in Fig.~\ref{fig:cz_chara}. Note that to keep error below $10^{-3}$, $\delta E_J/h$ can be allowed to vary within an $\sim\SI{60}{\mega\hertz}$ interval, while the variation of $J_{C}$ can be no more than $\sim\SI{0.5}{\mega\hertz}$. The variation of $\delta E_{J,T}$ mainly affects the phase error, while the leakage error varies around or below $10^{-5}$, thus is not shown in the plot. For the variation of $J_C$, on the other hand,  the leakage error dominates.

\section{The $i$SWAP Gate}\label{Sec:iswap}
The $i$SWAP gate is another type of two-qubit entangling gate that, combined with single qubit gates, is sufficient for universal quantum computation~\cite{Schuch_2003}. This gate is compromised by the parasitic $ZZ$ phase accumulation, and processes for eliminating such parasitic phases has been an important aspect for implementing high-fidelity $i$SWAP gate\cite{O_Malley_2015,McKay2016,Barends_2019}. The $i$SWAP gate requires a complete swapping process between $\ket{01}$ and $\ket{10}$, thus the modulation frequency $\omega_p$ should satisfy the resonance condition $\omega_p=\Delta$ of $\hat{H}_{i\mathrm{SWAP}}$ in Eq. \eqref{eq:intSWAP}. This requirement also means that the pulse should perform a complete $\pi$-rotation in the subspace spanned by $\ket{01}$ and $\ket{10}$, which implies  $g_{01,10}t_{\rm gate}\simeq \pi$, where $t_{\rm gate}$ is the gate time or the length of the pulse. Thus the range of pulse parameters is rather restricted. As a result, the extra phase on $\ket{11}$ state due to phase $\zeta$ accumulation, see Eq.~\eqref{eq:Uid}, cannot be easily compensated by optimization of the pulse. 

\begin{center}
    \begin{figure}
        \centering
        \includegraphics[width=0.8\textwidth]{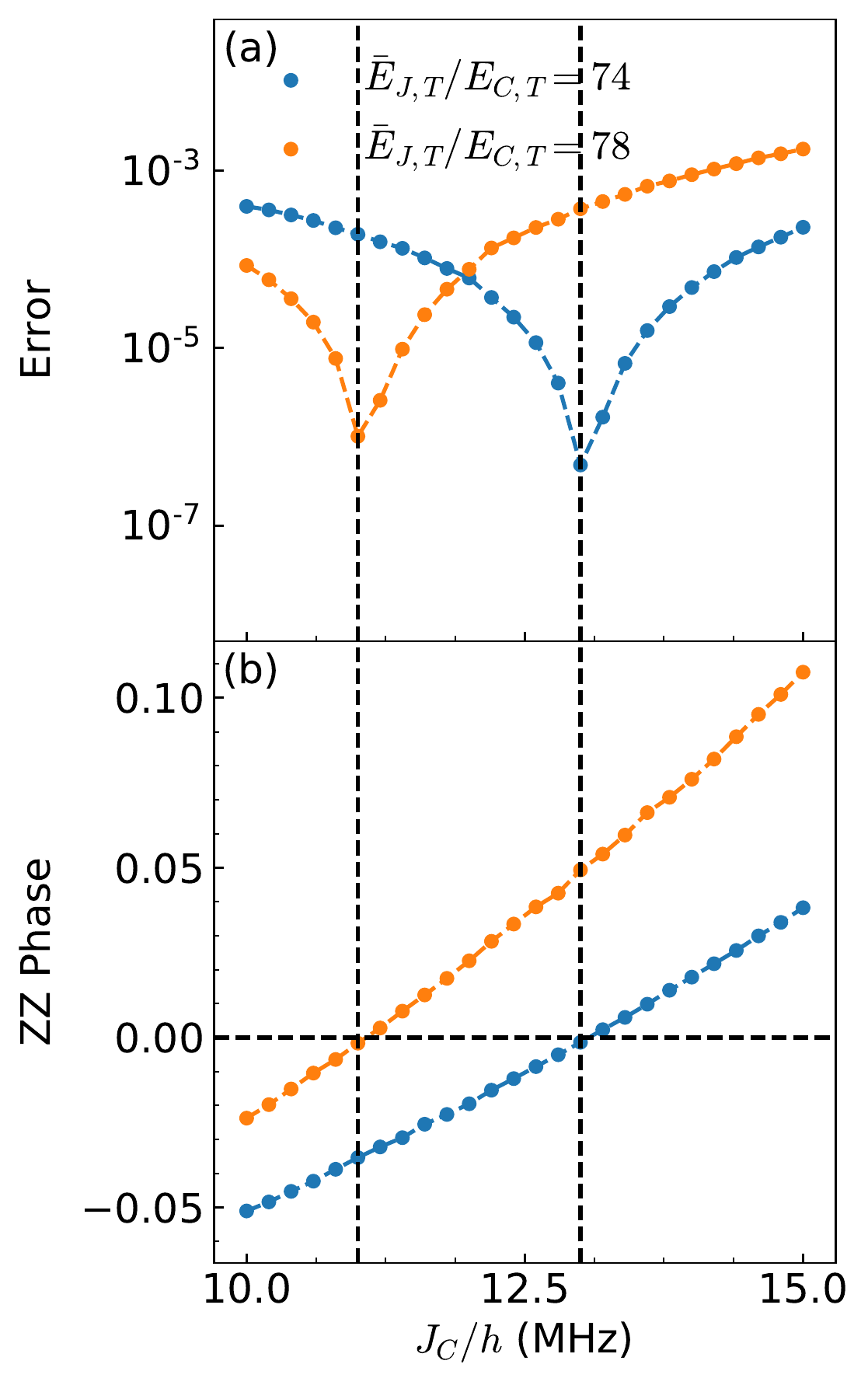}
        \caption{Simulation results for $i$SWAP gate for $E_{J,T}/E_{C,T}=74$ (blue) and $E_{J,T}/E_{C,T}=78$ (orange). (a) The total coherent error of $i$SWAP gate. (b) $ZZ$ phase accumulated during the $i$SWAP operation. The dashed lines are smooth fittings.}
        \label{fig:iswap_phase}
    \end{figure}
\end{center}

In Fig.~\ref{fig:iswap_phase} we show the simulation result for implementing an $i$SWAP gate with the same qubit parameters as in Sec.~\ref{Sec:cz}, except that we show the result for two different $\bar{E}_{J,T}$. As the top panel demonstrates, the infidelity curves have deep valleys for both cases but are clearly less than ideal when $J_C$ is away from the valley. The main contribution to the gate error is the phase error, as demonstrated by the bottom panel, where the $ZZ$ phase for the optimized gate is plotted against $J_C$. The valleys for fidelity curves happen right at the points where the $ZZ$ phase error crosses zero. As a numerical comparison, we model the error of incomplete rotation between $\ket{01}$ and $\ket{10}$ as $\hat{U}\ket{01}=\sin\gamma\ket{01}+i\cos\gamma\ket{10}$ and vice versa. The rotation error then again can be estimated by $2\sin^2(\gamma)/5$. The phase error, on the other hand, is modelled as an extra phase on $\ket{11}$ state $\hat{U}\ket{
11}=e^{i\delta\phi}\ket{11}$, and by Eq. \eqref{eq:fidelity}, it contributed approximately $3(\delta\phi)^2/20$ to the total error. Our numerical evaluation shows that the rotation error is several orders of magnitude less than the total error, except at the dip of the error curve, thus we do not show the rotation error in Fig.~\ref{fig:iswap_phase}. The phase error, on the other hand, is nearly indistinguishable from the total error, thus is also not displayed in the plot.

The unconditional  phase $\zeta$ mostly comes from the off-resonant state $\ket{00}$ and $\ket{11}$, $\zeta=\phi_{00}+\phi_{11}$. In Appendix \ref{App:phase}, we provide a perturbative estimate for such a phase, which captures the main contribution to $ZZ$ coupling. The corrections depend on the static energy spectrum and thus can be controlled by the static Josephson energy $\bar{E}_{J,T}$ of the tunable qubit. In Fig.~\ref{fig:iswap_phase}, we show the comparison of $i$SWAP gate for different $\bar{E}_{J,T}$ in the range of $\SI{10}{\mega\hertz}\leq J_C/h \leq \SI{15}{\mega\hertz}$. As demonstrated by the top panel, the infidelity curves now have the valleys at a different coupling $J_C$.  In the bottom panel, the change in $\bar{E}_{J,T}$ correspondingly induces changes in the slope and offset of the ZZ phase curve, thus also changing the point where the curve crosses zero. Therefore, a $ZZ$-free $i$SWAP gate can be achieved by tuning the static Josephson energy $\bar{E}_{J,T}$ of the tunable qubit, which can be implemented \emph{in situ} by tuning the gate voltage.

\begin{center}
    \begin{figure}
        \centering
        \includegraphics[width=\textwidth]{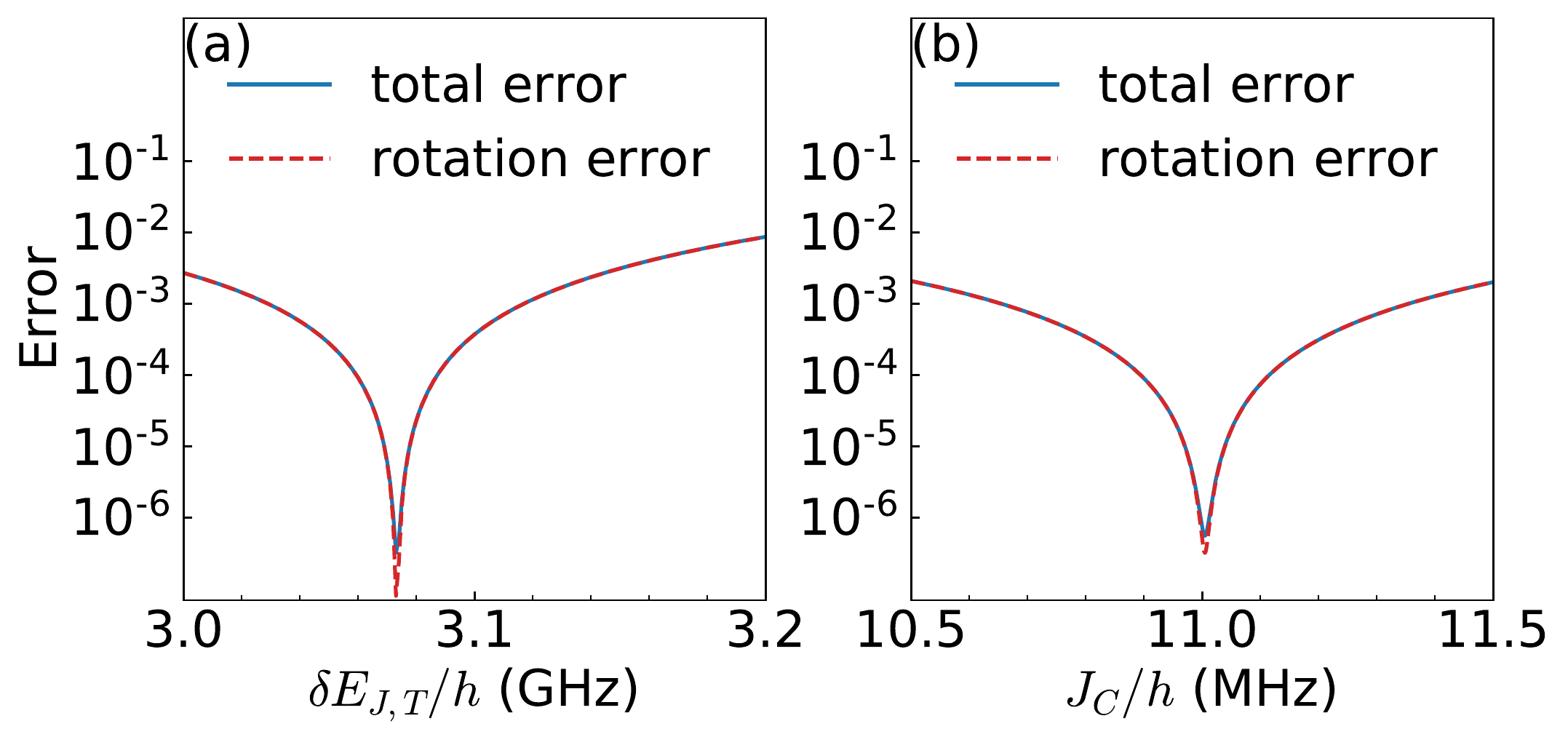}
        \caption{The total coherent error (blue solid) of $i$SWAP gate under the variation of (a) $\delta E_{J,T}$ and (b) $J_C$ , compared with the rotation error (red dashed). The gate parameters are optimized at $J_C/h=\SI{0.011}{\giga\hertz}$ for $\bar{E}_{J,T}/E_{C,T}=78$, as in Fig.~\ref{fig:iswap_phase}.}
        \label{fig:iswap_fluctuation}
    \end{figure}
\end{center}

In Fig.~\ref{fig:iswap_fluctuation} we again show the sensitivity of gate fidelity to the system parameters. The pulse parameters are optimized at $J_C/h=\SI{11}{\mega\hertz}$ for $\bar{E}_{J,T}/E_{C,T}=78$, as in Fig.~\ref{fig:iswap_phase}. The fidelity of $i$SWAP gate can be kept below $10^{-3}$ if the drive amplitude variates within a range of $\sim~\SI{60}{\mega\hertz}$, and if $J_C$ variation within of range of$~\sim\SI{0.6}{\mega\hertz}$. The gate error due to sensitivity is dominated by the rotation error. 


\section{The $\sqrt{i\mathrm{SWAP}}$ Gate} \label{Sec:sqrtiswap}

Here we provide simulations of $\sqrt{i\mathrm{SWAP}}$ gate. While the gate fidelity is also limited by an error due to an unwanted conditional phase $\zeta$,  for a $\sqrt{i\mathrm{SWAP}}$ gate, a complete swap between states $\ket{01}$ and $\ket{10}$ is no longer necessary as only a swap probability of 1/2 is needed. This loosens the restriction on the pulse parameters, and it is possible to correct the $ZZ$ phase by simply adjusting the pulse parameters. We find two ways to design the pulse shape: \emph{(i)} a one-tone pulse that performs an off-resonant Rabi oscillation and \emph{(ii)} a two-tone pulse that combines two sinusoidal pulses, with modulation frequency offset by some small value. For the error budget, we define the phase error is modeled and calculated in the same way as that for $i$SWAP gate, so it is estimated by $3(\delta\phi)^2/20$. We model the rotation error as $\hat{U}\ket{01}=\cos\left(\frac{\pi}{4}+\gamma\right)\ket{01}+i\sin\left(\frac{\pi}{4}+\gamma\right)\ket{10}$ where $\epsilon$ can be positive or negative, and similar error model holds for state $\ket{10}$. The rotation error is then approximately $3\sin^2(2\gamma)/20$. Since it is second-order effect, it is not the major error mechanism for $\sqrt{i\mathrm{SWAP}}$ gate.We also include the leakage error $\hat{U}\ket{11}=\cos\epsilon\ket{11}+\sin\epsilon\ket{02}$, which approximately contribute $9\sin^2(\epsilon)/40$ to the total error.

\subsection{One-tone pulses} 
A one-tone pulse achieves the $\sqrt{i\mathrm{SWAP}}$ gate by an off-resonant Rabi oscillation. To have a $\sqrt{i\mathrm{SWAP}}$ gate, the pulse parameter needs only to satisfy a weaker condition:
\begin{equation}\label{eq:swappingcondition}
    \dfrac{g_{01,10}}{\sqrt{g_{01,10}^2+\delta^2}}\sin(\dfrac{1}{2}\sqrt{g_{01,10}^2+\delta^2}t_{\mathrm{gate}})=\pm\dfrac{1}{\sqrt{2}}
\end{equation}
where $g_{01,10}$ is the effective coupling between $\ket{01}$ and $\ket{10}$ states, and $\delta=\omega_p-\Delta$ is the frequency detuning.  
This equation depends on the effective coupling $g_{01,10}$ and is not always satisfied. The off-resonant Rabi oscillation also introduces extra phases on the off-diagonal terms, but this does not change the $ZZ$ phase in the local-equivalent form, see Appendix~\ref{App:local_rot}. Nevertheless, the off-resonant case introduces change into the $g_{01,10}$ term, which then changes the offset of $ZZ$ phase accumulation rate as already explained for the $i$SWAP gate. Thus, it is possible to correct the $ZZ$ phase error for $\sqrt{i\mathrm{SWAP}}$ gate by only changing the pulse parameters. 


\begin{figure*}
    \centering
    \includegraphics[width=\textwidth]{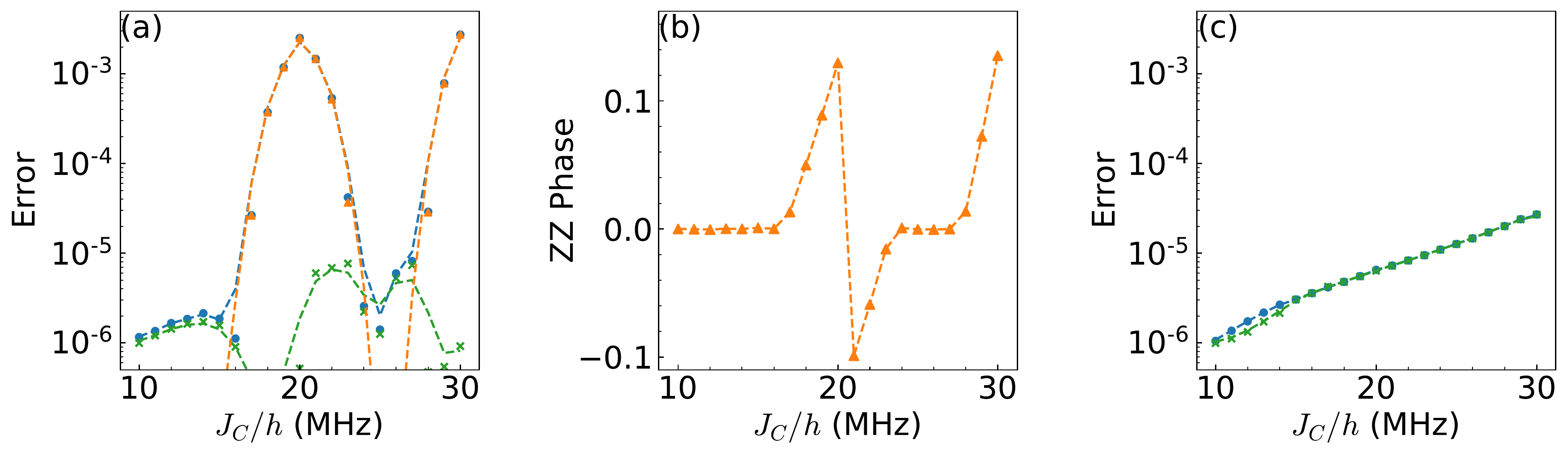}
\caption{The simulation results for one-tone and two-tone pulses $\sqrt{i\mathrm{SWAP}}$ gate. The dashed lines are smooth fittings. (a) The error for one-tone $\sqrt{i\mathrm{SWAP}}$ gate. The total error (blue dots) is decomposed into phase error (orange triangles) and leakage error (green crosses). (b) The conditional phase $\zeta$ of the one-tone gate is presented in (a).  The phase vanishes in a broad range of $J_C$, but cannot be reduced for other intervals of $J_C$, resulting in strong enhancement of the gate error in these intervals, cf. panel (a).  (c) The error for two-tone $\sqrt{i\mathrm{SWAP}}$ gate.  The total gate error (blue dots) remains small in the whole interval of $J_C$ and is dominated by leakage error (green crosses). The qubits parameters are the same as in Fig.~\ref{fig:cz_chara} or Fig.~\ref{fig:iswap_phase}. }
    \label{fig:sqrt_iswap_chara}
\end{figure*}

Fig.~\ref{fig:sqrt_iswap_chara}(a) shows the simulation result of $\sqrt{i\mathrm{SWAP}}$ gates using one-tone pulses, with coupling $J_C$ ranging from $\SI{10}{\mega\hertz}$ to $\SI{30}{\mega\hertz}$. The $ZZ$ phase curve now has a plateau of zero phase error, rather than a single point intersection with the $J_C/h$-axis, as shown by Fig.~\ref{fig:sqrt_iswap_chara}(b). The endpoints of the zero plateau correspond to two types of extrema for the $ZZ$ phase, one happens close to resonance, and the other is related to the parameter $g_{01,10}$. In the plateau region, the phase error is insignificant, and the error is dominated by the leakage error out of the computational subspace. On the other hand, in the parameter range outside the plateau, the $ZZ$ phase deviates from zero, and the error curve has peaks. The error budget shows that the major contribution of error along the peaks comes from the phase error. Also, there is a discontinuity of $ZZ$ phase error at $J_C=\SI{20}{\mega\hertz}$, which corresponds to the discontinuity in the pulse parameters. This discontinuity is related to the choice of a different branch of the Rabi-oscillation period when the swapping process switches from a $3\pi/2$ rotation to a $5\pi/2$ rotation; the detail of the analysis above can be found in Appendix \ref{App:branch}.

\subsection{Two-tone pulses} 
Another way to correct the $ZZ$ phase error is to composite two pulses with close but different modulation frequencies, so that the combined pulse is
\begin{equation}
\begin{aligned}
    E_{J,T}(t)&=\bar{E}_{J,T}\\&+
    f(t)(\delta E_{J,T,1}\cos(\omega_{p,1}t)+\delta E_{J,T,2}\cos(\omega_{p,2}t))\,.
\end{aligned}
\end{equation}
This introduces new pulse parameters, adding more degree of freedom for $ZZ$ phase elimination. As shown in Fig.~\ref{fig:sqrt_iswap_chara}(c) for $J_C/h$ ranging form $\SI{10}{\mega\hertz}$ to $\SI{30}{\mega\hertz}$, the two-frequency method can correct the $ZZ$-phase error more efficiently, restricting the $ZZ$ phase error to have amplitude $<10^{-3}$. We only optimize the driving amplitudes of the two-tone pulses, but keep modulation frequencies $\omega_{p,1}$ and $\omega_{p,2}$ fixed across different $J_C$. The reason for this is to simplify the optimization procedure, but as Fig.~\ref{fig:sqrt_iswap_chara}(c) shows, although the infidelity increases monotonically with $J_C$, it can be kept $<10^{-4}$ across this range of $J_C$, thus our way to only optimize driving amplitudes can still be justified. The monotonic increase is mainly the result of increasing leakage of $\ket{11}$ state, which increases from the order of $10^{-6}$ to the order of $10^{-4}$ as $J_C$ increases from $\SI{10}{\mega\hertz}$ to $\SI{30}{\mega\hertz}$, while the other error mechanisms, including the phase error, contribute below $10^{-7}$ to the total error, several orders of magnitude less than the leakage error. This increase in leakage is due to the increasing coupling constant $J_C$, which also increases the coupling between $\ket{11}$ and $\ket{02}$ state.

\section{Effect of Decoherence on Gates}\label{Sec:decoherence}

The results in the previous sections were obtained for unitary system evolution. For physical qubits, we also need to take the decoherence process into consideration. Therefore, we also use the same qubit and pulse parameters to simulate the following master equation with finite relaxation times,
\begin{equation}
    \begin{aligned}
    \Dot{\hat{\rho}}(t)&=-\dfrac{i}{\hbar}\comm{\hat{H}(t)}{\hat{\rho}(t)}+\sum_{\alpha=F,T}\dfrac{1}{T_{1,\alpha}}\mathcal{D}_{\alpha}(\hat{\rho}(t))\,.
    \end{aligned}
\end{equation}
Here $\hat{H}(t)$ is the two-qubit Hamiltonian defined as in Eq.~\eqref{eq:2qbit}, and the superoperator $\mathcal{D}_{\alpha}$ is defined as 
\begin{equation}
\begin{aligned}
    \mathcal{D}_{\alpha}(\hat{\rho})=&\sum_{j=0}^{j_{t}} 2\hat{c}_{\alpha,j}\hat{\rho}\hat{c}_{\alpha,j}^{\dagger}-\hat{\rho}\hat{c}_{\alpha,j}^{\dagger}\hat{c}_{\alpha,j}-\hat{c}_{\alpha,j}^{\dagger}\hat{c}_{\alpha,j}\hat{\rho}\\
    \hat{c}_{F,j}=&\sqrt{j+1}\sum_{i}\ket{j,i}\bra{j+1,i}\\ \hat{c}_{T,j}=&\sqrt{j+1}\sum_{i}\ket{i,j}\bra{i,j+1}\,.
\end{aligned}
\end{equation}
The $T_{1,\alpha}$ stands for the relaxation times of each qubit, and for convenience, we will set $T_{1,F}=T_{1,T}$ in simulations. Here $j_t$ is the truncation number of states. For $t_{\rm gate}\ll T_1$, we expect that the fidelity $F$ will follow an exponential relation with respect to the gate time, so $1-F\simeq 4t_{\rm gate}/5T_1$ (assuming that $T_2=2T_1$)~\cite{Tripathi_2019}. For example, the error for both gates around $T_1=\SI{100}{\micro\second}$ is around $6\times 10^{-4}$, close to the theoretical prediction. The simulated fidelity $F$ is calculated with process fidelity $F_p$ and chi matrix $\chi$ of the process by $F=[4F_p+\Tr(\chi)]/5$~\cite{Nielsen_2002}.

\begin{center}
    \begin{figure}
        \centering
        \includegraphics[width=0.7\textwidth]{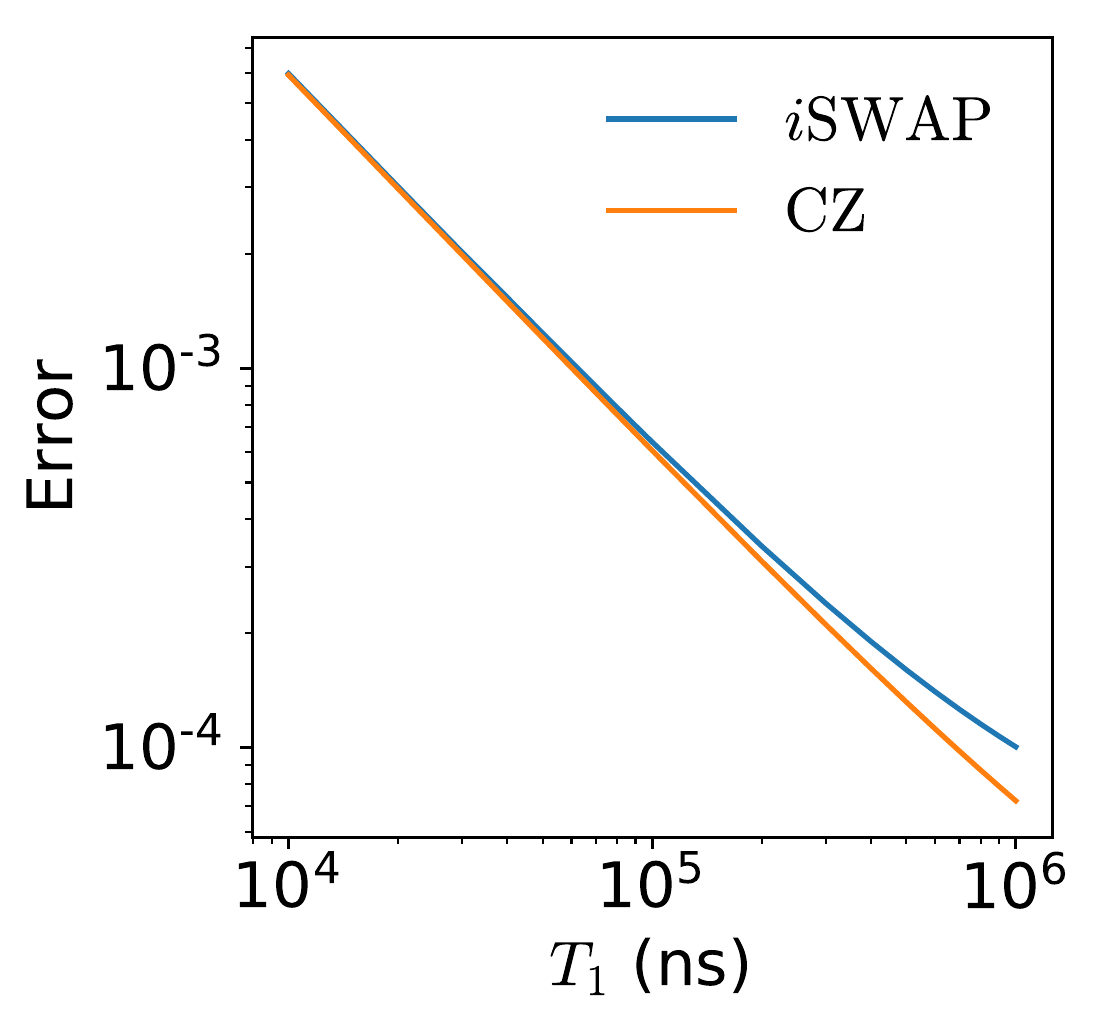}
        \caption{Simulations with dissipation for CZ  and $i\mathrm{SWAP}$ gates. The qubit and pulse parameters  are the same as in Fig.~\ref{fig:cz_chara} and Fig.~\ref{fig:iswap_phase}, with $J_c/h=\SI{10}{\mega\hertz}$. As $T_1$ approaches $\SI{1}{\milli\second}$, the gate error for $i$SWAP saturates towards its unitary estimate of $\sim 8\times 10^{-5}$, which is higher than that estimate for the CZ gate ($\sim 10^{-6}$).
} 
        \label{fig:dissipation}
    \end{figure}
\end{center}

In Fig.~\ref{fig:dissipation} we show the simulation results with dissipation for both CZ gate and $i$SWAP gate, using parameters for Figs.~\ref{fig:cz_chara} and ~\ref{fig:iswap_phase} at $J_C/h=\SI{10}{\mega\hertz}$. We also simulate for both one-tone and two-tone $\sqrt{i\mathrm{SWAP}}$ gate, which is not shown in the plot because the curves of $\sqrt{i\mathrm{SWAP}}$ gate fidelity and $i$SWAP gate fidelity as a function of $T_1$ are nearly 
indistinguishable on the plot. We note that for fault-tolerant error correction code, the plot suggests that $T_1$ should be greater than $\SI{10}{\micro\second}$ to achieve the reported threshold error rate $0.57\%$ for surface code~\cite{Fowler_2012}. In truth, several experiments that use typical transmon qubits to implement surface code report $T_1$'s that are above this threshold~\cite{Varbanov_2020,Alibaba_2020}. However, if we aim for higher fidelity that surpasses $0.1\%$, we need at least $T_1\gtrsim \SI{70}{\micro\second}$. Recent nanowire gatemons have seen an improvement of $T_1$ to over $\SI{20}{\micro\second}$~\cite{Luthi_2018}, which would satisfy the former threshold, but still needs to be further improved to pass the latter. Therefore, understanding the decoherence mechanism and improving the qubit coherence should be a crucial part for development of gatemon-based quantum processors.

\section{Conclusion}

We described implementation of parametric entangling gates on superconductor-semiconductor hybrid qubits using voltage modulation pulses. We simulate CZ gate, $i$SWAP gate, $\sqrt{i\mathrm{SWAP}}$ with this idea, and demonstrate that all gates can have coherent error $<10^{-5}$, with proper choice of qubit and pulse parameters. In particular, the CZ gate has the least stringent conditions and does not require an extra procedure to correct the accumulated $ZZ$ phase. The $i$SWAP gate, on the other hand, is mainly limited by the $ZZ$ phase but can be corrected by tuning the static Josephson energy, which changes the offset of the $ZZ$ phase curve by an ac-Stark shift term. The $\sqrt{i\mathrm{SWAP}}$ has less stringent conditions than the $i$SWAP gate, thus can be partially corrected by choice of pulse parameters alone. Using one-tone pulses, the conditional $ZZ$ phase can be corrected in a finite range, unlike that in the $i$SWAP case where the correctable static $ZZ$ phase has to be equal to the offset determined by the qubit parameters. For two-tone pulses, the $ZZ$ phase accumulation can be corrected in a wider range of qubit parameters by tuning the pulse amplitudes only, in this case, the gate performance is limited by leakage errors. 

We also simulate the sensitivity of the gate fidelity to the variations of the drive amplitude and the coupling constant. We find that to keep the coherent error below $10^{-3}$, the variation of $\delta E_{J,T}$ should be no more than $\SI{60}{\mega\hertz}$ for both CZ and $i$SWAP gate, while the variation of $J_C$ should be no more than $\SI{0.5}{\mega\hertz}$ for CZ gate, and no more than $\SI{0.6}{\mega\hertz}$ for $i$SWAP gate. For decoherence, we find that to have fault-tolerant gates ready for error correction, we need $T_1>\SI{70}{\micro\second}$. The current state-of-art superconductor-semiconductor hybrid qubits fall short of this demand, but our simulation may set a standard for future improvements in hardware.

Although the coherence of gatemon qubit is still less than ideal, future advancement in hardware fabrication can alleviate such limit and extend gatemon's $T_1$ to be above the threshold mentioned earlier. Further investigation of superconducting-semiconducting hybrid structures could bring fruitful results for developing voltage-controlled superconducting circuits, such as gatemon qubits and tunable resonators~\cite{Strickland_2022}.

\begin{acknowledgments}
The authors acknowledge support from the Army Research Office agreement W911NF2110303. We performed computations using resources and assistance of the UW-Madison Center For High Throughput Computing (CHTC) in the Department of Computer Sciences. The CHTC is supported by UW-Madison, the Advanced Computing Initiative, the Wisconsin Alumni Research Foundation, the Wisconsin Institutes for Discovery, and the National Science Foundation.
\end{acknowledgments}

\appendix
\section{Interaction Hamiltonian}\label{App:int}

To derive a more complete Hamiltonian, we follow the method presented in Ref~\cite{Didier2018}, but make changes to fit our model. Define the variable $\xi_i=\sqrt{2E_{C,i}/E_{J,i}}$, where $i=F,T$. With the transformation into instantaneous eigenbasis of single qubit, we may write the truncated Hamiltonian as~\cite{Didier2018},
\begin{widetext}
\begin{equation}
\begin{aligned}
\hat{H}_{M}(t)=&\sum_{i=0}^{2}[\nu_{F,i}\hat{\Pi}_{F,i}+\nu_{T,i}(t)\hat{\Pi}_{T,i}]\\
+&g(t)[\lambda_{F}\hat{\sigma}_{F,01}^y+\sqrt{2}\Lambda_{F}\hat{\sigma}_{F,12}^y][\lambda_{T}(t)\hat{\sigma}_{T,01}^y+\sqrt{2}\Lambda_{T}(t)\hat{\sigma}_{T,12}^y]\\
+&\mu(t)\hat{\sigma}_{T,02}^y\,,
\end{aligned}
\end{equation}
\end{widetext}
where $\hat{\Pi}_{n,i}$ is projection operator of $n$-th single qubit eigenstate, and $\hat{\sigma}_{Q,ij}^y$ stands for the Pauli $y$ matrix between single qubit states $\ket{i}$ and $\ket{j}$ for qubit $Q$. The coefficients $\lambda_i$ and $\Lambda_i$ are determined by $\xi_i$, but in the transmon regime, they are close to unity. The notation $\nu_{Q,i}(t)$ stands for the instantaneous frequencies of qubit $Q$, and we use 
\begin{equation}
    \omega_{Q,ij}(t)=\nu_{Q,j}-\nu_{Q,i}
\end{equation}
to denote the single qubit instantaneous transition frequency. The coupling $g$ is defined by,
\begin{equation}
    g(t)=J_C/(4\sqrt{\xi_F\xi_T(t)})\,,
\end{equation}
and
\begin{equation}
    \mu(t)=\bra{0_{T}(t)}\dfrac{d}{dt}\ket{2_{T}(t)}\,.
\end{equation}
Here $\ket{i(t)}$ is the $i$-th eigenstate for the time-dependent Hamiltonian. The $0$ to $2$ transition term is present in the interaction picture because of the $\hat{\varphi}^2$-perturbation after expanding the $\cos$ term in the single qubit Hamiltonian. Also, for the combined system, we use the notation,
\begin{equation}
    \nu_{ij}(t)=\nu_{F,i}+\nu_{T,j}(t)\,,
\end{equation}
and denote the two-qubit transition frequencies as
\begin{equation}
    \omega_{ij,rs}(t)\equiv \nu_{rs}(t)-\nu_{ij}(t)\,.
\end{equation}

Now assume a periodic modulation with frequency $\omega_p$ on the tunable qubit. This means that we can write down the instantaneous eigenfrequencies for the tunable qubit as a Fourier series
\begin{equation}
    \nu_{T,i}(t)=\sum_{m=0}^{\infty}\nu_{m;T,i}\cos(m\omega_pt)\,,
\end{equation}
and similarly for $\omega_{T,ij}(t)$, $\nu_{ij}(t)$ and $\omega_{ij,rs}(t)$. Before transform to interaction picture, we first need the Jacobi-Auger identity, which states that $e^{ix\sin y}=\sum_{l=-\infty}^{\infty} J_n(x)e^{iny}$, where $J_n$ is the Bessel function of first kind. Then:
\begin{equation}
\begin{aligned}
    e^{i\int_{0}^{t}dt'\omega_{ij,rs}(t)}&=e^{if_{0}t}\sum_{n\in\mathbb{Z}}G_{n}(\{\omega_{k;rs,ij}\})e^{in\omega_{p}t}\\ \{\omega_{k;rs,ij}\}&=(\omega_{0;rs,ij},\omega_{1;rs,ij},\cdots)\,,
\end{aligned}
\end{equation}
the coefficient $G_n$ is
\begin{equation}\label{eq:k_mode}
    G_{n}(\{\omega_{k;rs,ij}\})=\sum_{(l_k)\in S_n}\left(\prod_{k=1}^{\infty}J_{l_{k}}\left(\dfrac{\omega_{k;rs,ij}}{k\omega_{p}}\right)\right)
\end{equation}
where the set $S_n$ of integer sequences is defined as:
\begin{equation}
    S_n=\{(l_1,l_2,\cdots)|l_k\in\mathbb{Z},\sum_{k=1}^{\infty} kl_k=n\}
\end{equation}

Now transform the Hamiltonian $\hat{H}_M$ into the interaction picture with the unitary matrix
\begin{equation}\label{eq:intpic}
\hat{U}_{int}=\exp{-i\int_{0}^{t}dt'\sum_{i=0}^{2}[\nu_{F,i}\Pi_{F,i}+\nu_{T,i}\Pi_{T,i}]}\,,
\end{equation}
which eliminates the diagonal element. With the notation above, the matrix elements then transform as:
\begin{widetext}
\begin{equation}
    \hat{U}_{\rm int}\ket{ij}\bra{rs}\hat{U}_{\rm int}^{\dagger}=e^{i(\omega_{0,rs}-\omega_{0,ij})t}\sum_{n\in\mathbb{Z}}G_{n}(\{\omega_{k;rs,ij}\})e^{in\omega_{p}t}\ket{ij}\bra{rs}
\end{equation}
From the off-diagonal terms in $H_M(t)$, we expect that
\begin{equation}\label{eq:full_int}
\begin{aligned}
H_{\rm int}(t)=&\sum_{n\in\mathbb{Z}}\sum_{ij,rs}g_{ij,rs}^{(n)}(t)e^{i(\omega_{0,rs}-\omega_{0,ij}+n\omega_{p})t}
+\sum_{n\in\mathbb{Z}}\Omega_{02}^{(n)}(t)e^{i(n\omega_p+2\bar{\omega}_{T,01}-\bar{\eta}_T)t}I_3\otimes\ket{2}\bra{0}+\mathrm{h.c.}
\end{aligned}
\end{equation}
where $n$ runs over all the integers. The coupling coefficients and resonant frequencies for are:
\begin{equation}\label{eq:coupling} 
\begin{aligned}
    g_{00,11}^{(n)}(t)&=-g(t)\lambda_F\lambda_T(t)G_{n}(\{\omega_{k;11,00}\}),\qquad &\omega_{0;11,00}&=\omega_{F,01}+\bar{\omega}_{T,01};\\
    g_{01,10}^{(n)}(t)&=g(t)\lambda_F\lambda_T(t)G_{n}(\{\omega_{k;10,01}\}),\qquad  &\omega_{0;10,01}&=-\omega_{F,01}+\bar{\omega}_{T,01};\\
    g_{11,22}^{(n)}(t)&=-2g(t)\Lambda_F\Lambda_T(t)G_{n}(\{\omega_{k;22,11}\}),\qquad &\omega_{0;22,11}&=\omega_{F,01}+\bar{\omega}_{T,01}-\eta_F-\bar{\eta}_T;\\
    g_{12,21}^{(n)}(t)&=2g(t)\Lambda_F\Lambda_T(t)G_{n}(\{\omega_{k;21,12}\}),\qquad &\omega_{0;21,12}&=\omega_{F,01}-\bar{\omega}_{T,01}-\eta_F+\bar{\eta}_T;\\
    g_{01,12}^{(n)}(t)&=-\sqrt{2}g(t)\lambda_F\Lambda_T(t)G_{n}(\{\omega_{k;12,01}\}),\qquad &\omega_{0;12,01}&=\omega_{F,01}+\bar{\omega}_{T,01}-\bar{\eta}_T;\\
    g_{02,11}^{(n)}(t)&=\sqrt{2}g(t)\lambda_F\Lambda_T(t)G_{n}(\{\omega_{k;11,02}\}),\qquad &\omega_{0;11,02}&=\omega_{F,01}-\bar{\omega}_{T,01}+\bar{\eta}_T;\\
    g_{10,21}^{(n)}(t)&=-\sqrt{2}g(t)\Lambda_F\lambda_T(t)G_{n}(\{\omega_{k;21,10}\}),\qquad &\omega_{0;21,10}&=\omega_{F,01}+\bar{\omega}_{T,01}-\eta_F;\\
    g_{11,20}^{(n)}(t)&=\sqrt{2}g(t)\Lambda_F\lambda_T(t)G_{n}(\{\omega_{k;20,11}\}),\qquad &\omega_{0;20,11}&=\omega_{F,01}-\bar{\omega}_{T,01}-\eta_F\,;
\end{aligned}
\end{equation}
and
\begin{equation}
    \Omega_{02}^{(n)}(t)=\mu(t)G_{n}(\{\omega_{k;T,02}\})\,, \qquad \omega_{0;T,02}=2\bar{\omega}_{T,01}-\bar{\eta}_T.
\end{equation}
\end{widetext}
Although the expression for $G_n$ complicated, for weak modulation on the system spectrum, the leading contribution to $G_n$ can be simplified to the Bessel function $J_n(\omega_{1;ij,i'j'}/\omega_p)$, where $\omega_{1;ij,i'j'}$ is the amplitude of frequency modulation to the harmonic mode, which is determined by the Josephson modulation amplitude in the actual system. In particular, for $n=0$:
\begin{equation}
    G_0(\{\omega_{k;ij,i'j'}\})\approx J_0\left(\dfrac{\omega_{1;ij,i'j'}}{\omega_p}\right)\,,
\end{equation}
and for $n=1$,
\begin{equation}
    G_1(\{\omega_{k;ij,i'j'}\})\approx J_1\left(\dfrac{\omega_{1;ij,i'j'}}{\omega_p}\right)\,.
\end{equation}
When the modulation is weak, we can further simplify by the approximation $J_0(\omega_{1;ij,i'j'}/\omega_p)\approx 1$ and $J_1(\omega_{1;ij,i'j'}/\omega_p)\approx \omega_{1;ij,i'j'}/2\omega_p$. For $g(t)$ that is nearly constant under weak modulation, the $n=0$ terms then roughly correspond to the static Hamiltonian, and the $n=1$ terms correspond to the drive Hamiltonian as presented in Eqs.~\eqref{eq:int}.

For the qualitative analysis below, we will assume $g^{(n)}$ to be constant for convenience.

\section{Qualitative Analysis on Phase Accumulation for off-resonant States}\label{App:phase}

The unitary operator used in \eqref{eq:intpic} is essentially a single qubit operation. Therefore, to analyze the $ZZ$ phase due to the evolution, it is sufficient to look at the interaction Hamiltonian \eqref{eq:full_int}. For convenience, let us denote the two qubit state by a single letter $k$, $l$, etc. We can write the relevant two-qubit term as,
\begin{equation}
    H_{r}=\sum_n\sum_{k\neq l}g^{(n)}_{kl}\ket{k}\bra{l}e^{i\omega^{(n)}_{kl}t}\,,
\end{equation}
with the summation of $n$ runs over all integers. Here we assume $g^{(n)}_{kl}=g^{(-n)}_{kl}$ are real effective coupling between states $\ket{k}$ and $\ket{l}$ for each mode $n$, and $\omega^{(n)}_{kl}=-\omega^{(-n)}_{kl}=\omega_{0;l}-\omega_{0;k}+n\omega_p$, with notation as in Appendix \ref{App:int}. Thus we have the relation: 
\begin{equation}
    \omega^{(n)}_{lr}+\omega^{(n')}_{rk}=\omega^{(n+n')}_{kl}\,.
\end{equation}

Suppose the system state is $\psi(t)=\sum c_k\ket{k}$. The time evolution of the state then satisfies the first-order time differential equation:
\begin{equation}
    \dot{c}_l=\sum_{n\in\mathbb{Z}}\sum_k -ig^{(n)}_{lk}e^{i\omega^{(n)}_{lk}t}c_k.
\end{equation}
This linear differential equation is not easy to solve. However, since the Hamiltonian is bounded and Lipschitz, it is reasonable to assume that the recursion relation:
\begin{equation}
    \psi^{(i+1)}(t)=\psi(0)-i\int_0^tH_r(t')\psi^{(i)}(t')dt'
\end{equation}
converges to a fixed point~\cite{viana2021differential}. With RWA, by choosing a proper $\psi^{(0)}$, we can assume that the recursion converges rapidly, thus giving us a good estimate within a few recursive steps.

Now assume that the initial state is $\psi(0)=\ket{k}$, such that no $\omega_{kl}^{(n)}$ is close to 0. Suppose also that $\abs{\omega_{ll'}^{(n)}}\gg \abs{g_{ll'}}$ for all $l,l'$ in concerns, and the terms with $\omega_{ll'}^{(n)}\approx 0$ do not contribute. In this case, we expect that $\psi(t)$ does not deviate from $\ket{k}$ too much by the RWA assumption. Thus we have $c_k^{(0)}=1$, and

\begin{widetext}
\begin{equation}
\begin{aligned}
    c_{l}^{(2)}&=-\sum_{n\in\mathbb{Z}}\dfrac{g_{lk}^{(n)}}{\omega_{lk}^{(n)}}(e^{i\omega_{lk}^{(n)}t}-1)\\&+\sum_{n,n'\in\mathbb{Z}}\sum_{l'\neq l,k}\dfrac{g_{l'k}^{(n)}g_{ll'}^{(l)}}{\omega_{l'k}^{(n)}}\left(\dfrac{1}{\omega_{ll'}^{(n+n')}}(e^{i\omega_{kl'}^{(n+n')}t}-1)-\dfrac{1}{\omega_{ll'}^{(l)}}(e^{i\omega_{ll'}^{(l)}}-1)\right)\quad l\neq k\\
    c_{k}^{(2)}&=1+\sum_{n\in\mathbb{Z}}\sum_{k\neq l}\dfrac{\abs{g_{lk}^{(n)}}^{2}}{\omega_{lk}^{(n)}}it+\sum_{n\neq n'\in\mathbb{Z}}\sum_{k\neq l}\dfrac{g_{lk}^{(n)}g_{lk}^{(n')}}{\omega_{lk}^{(n)}(n-n')\omega_{p}}(e^{i(n-n')\omega_{p}t}-1)\\&+\sum_{n,n'\in\mathbb{Z}}\sum_{k\neq l}\dfrac{g_{lk}^{(n)}g_{kl}^{(n')}}{\omega_{lk}^{(n)}\omega_{kl}^{(n')}}(e^{i\omega_{kl}^{(n')}t}-1)
    \,.
\end{aligned}
\end{equation}
\end{widetext}

Let 
\begin{equation}
\epsilon=\max_{n,l\neq l'}\left\{\abs{\dfrac{g^{(n)}_{l,l'}}{\omega^{(n)}_{l,l'}}},\dfrac{g_{l,l'}^{(n)}}{\omega_p}\right\}\,,
\end{equation}
the largest correction then is present in the $\abs{c_n^{(2)}}$ term as linear in time $t$. Thus, to the leading order of $\epsilon$, the phase accumulation rate of $c_k$ is approximately:
\begin{equation}
    \phi_k/t\approx\sum_{n}\sum_{l}\dfrac{\abs{g^{(n)}_{lk}}^2}{\omega^{(n)}_{lk}}\,.
\end{equation}

\begin{table}
    \centering
\begin{tabular}{c|c}
    \hline\hline
     Transition Process & Transition Frequency (GHz)\\
    \hline
    $00\leftrightarrow 11$ & 10.1058\\
    $01\leftrightarrow 10$ & 0.7923\\
    $11\leftrightarrow 22$ & 9.6635\\
    $12\leftrightarrow 21$ & 0.7959\\
    $02\leftrightarrow 11$ & 1.0151\\
    $10\leftrightarrow 21$ & 9.8863\\
    $11\leftrightarrow 20$ & 0.5734\\
    \hline\hline
\end{tabular}
\caption{Relevant transtion frequencies of the two-qubit system, calculated at $J_C=\SI{10}{\mega\hertz}$.}
\label{tab:freq}
\end{table}

We note that for such approximation to hold, $g_{l,l'}^{(n)}$ should be consistenly smaller than $\omega_{l,l'}^{(n)}$ when there is no resonance between $m$ and $m'$. We argue from Eq.~\eqref{eq:k_mode} that $g_{l,l'}^{(n)}$ has a leading term $J_m(\omega_{1;l,k})_1/\omega_p)$, which asymptotically is restricted by $\frac{1}{m!}(\frac{\omega_{1;l,k}}{2\omega_p})^m$ for $\omega_{1;l,k}/\omega_p\ll\sqrt{m+1}$~\cite{abramowitz_stegun}.
Therefore, we expect the contribution from higher $m$-modes to be small enough, and only focus on the order of magnitudes for lower $m$-modes. In particular, we are using $m=1$ mode to build the gate, so the main concern is for the $k=0$ and $k=1$ modes. For concreteness, in Table~\ref{tab:freq} we show an example of relevant transition frequencies for the chosen qubit parameters. The qubit parameters are the same as those used for simulation, with $J_C=\SI{10}{\mega\hertz}$. Using the static spectrum as a reference, we expect that for $m=0$ mode, $\omega_{m;l,k}$ has the lower bound to be about $\SI{0.5}{\giga\hertz}$. For $m=1$ mode, we are concerned with the magnitude of $\omega_{m;l,k}-\omega_p$, where $\omega_p$ should be close to transition frequency of either $01\leftrightarrow 10$ or $11\leftrightarrow 02$. The lower bound is the difference between the transition frequency between $01\leftrightarrow 10$ and $12\leftrightarrow 21$, which is about $\SI{0.003}{\giga\hertz}$, but since the states $\ket{12}$ and $\ket{21}$ are out of computational subspace, and there is no direct coupling that relates these two transition processes, we argue that this lower bound should not have a significant contribution to the correction. The next lower bound is the difference between transition frequencies of $01\leftrightarrow 10$ and $11\leftrightarrow 02$ or $11\leftrightarrow 20$, which is about $\SI{0.2}{\giga\hertz}$. As a comparison, $g_{mn}^{(1)}$ should be of the order of $\pi/t_{\rm gate}$, which for our choice of gate time is about $\SI{0.04}{\giga\hertz}$. Thus, we argue that the approximation we make here should be reasonable for a rough estimate.

The $m=0$ term is approximately the static $ZZ$ interaction rate found by second-order perturbation theory, and determines the slope in Fig.~\ref{fig:iswap_phase}(b). The $m=1$ terms provide the $ZZ$-phase a downward shift, and are similar to the ac Stark shift that is present in the off-resonant microwave method~\cite{Chow_2013,Paik_2016,Krinner_2020,Xiong_2022}. The difference is that in the microwave drive method, the numerator is the amplitude of the transverse drive, while here the numerator is given by \eqref{eq:coupling}, which has a more complicated dependence on the pulse shape and modulation amplitude. As the $i$SWAP gate requires that $g^{(1)}_{01,10}=\pi/t_{\rm gate}$, and all the other $g_{kl}^{(1)}$ should be proportional to $g_{01,10}^{(1)}$ as they correspond to the same harmonic mode, we see that the offset is then approximately:
\begin{equation}
    \sum_{k,l}\dfrac{\pi^2}{\omega^{(1)}_{k,l} t_{\rm gate}}\abs{\dfrac{g_{k,l}^{(1)}}{g_{01,10}^{(1)}}}^2+\sum_{k,l}\dfrac{\pi^2}{\omega^{(-1)}_{k,l} t_{\rm gate}}\abs{\dfrac{g_{k,l}^{(1)}}{g_{01,10}^{(1)}}}^2
\end{equation}

\begin{figure}
    \centering
    \includegraphics[width=0.8\textwidth]{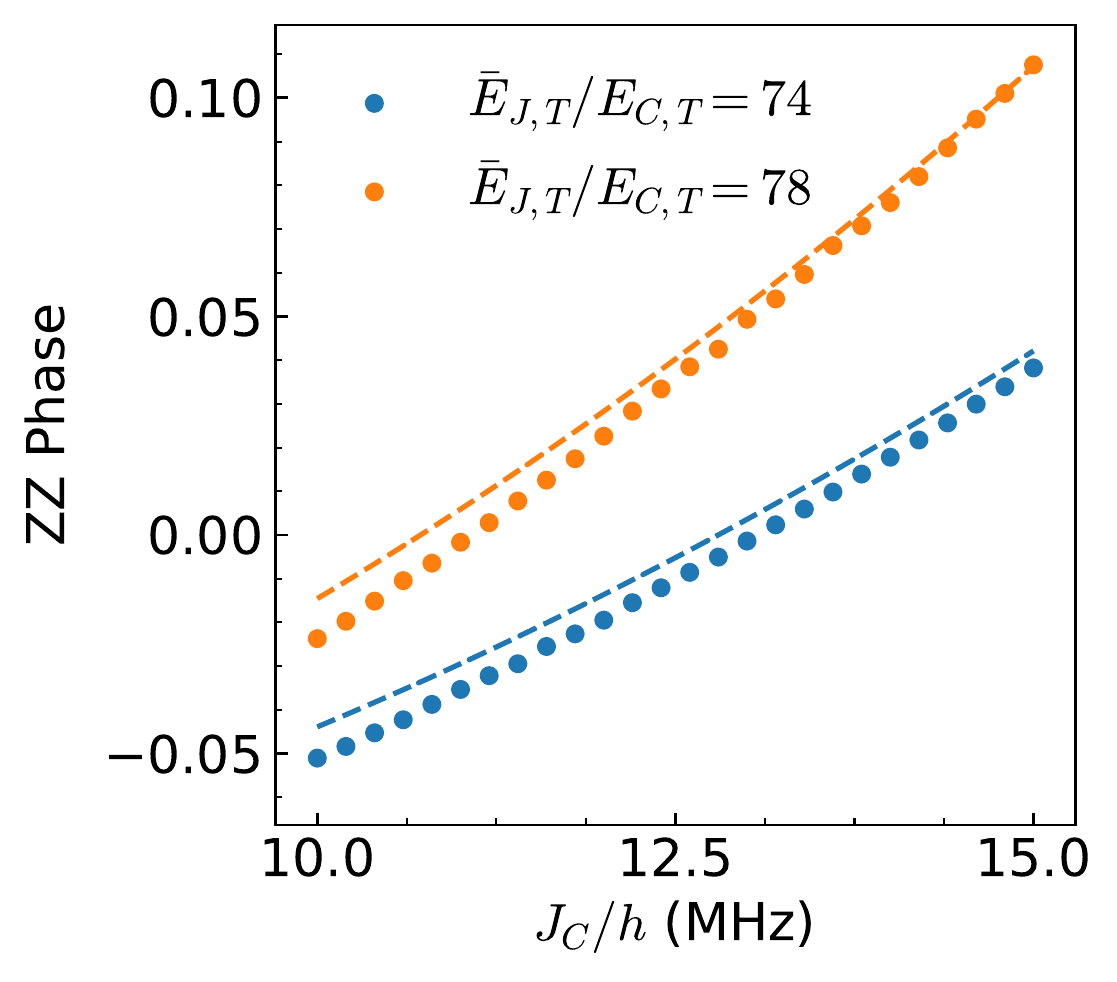}
    \caption{Comparison between the theoretical estimation and the numerical simulation as presented in Fig.~\ref{fig:iswap_phase}(b). The dots are the numerical value, and the dashed lines are the theoretical estimation. }
    \label{fig:phase}
\end{figure}

In summary, for off-resonant state $\ket{ij}$, the accumulation rate of $\phi_{ij}$ as determined by $n=0,1$ harmonic mode is

\begin{widetext}
\begin{equation}\label{eq:zz_rate}
    \dfrac{\phi_{ij}}{t_{\rm gate}}\approx\sum_{kl\neq ij}\dfrac{(g^{(0)}_{ij,kl})^2}{\bar{\omega}_{kl}-\bar{\omega}_{ij}}+\sum_{kl\neq ij}\dfrac{2(\bar{\omega}_{kl}-\bar{\omega}_{ij})}{(\bar{\omega}_{kl}-\bar{\omega}_{ij})^2-\omega_p^2}(g^{(1)}_{ij,kl})^2\,,
\end{equation}
\end{widetext}

where $\bar{\omega}_{ij}$ stands for the time average of corresponding frequency. As a demonstration, we compare the theoretical estimation and numerical result in Fig.~\ref{fig:phase}. We use static $g^{(0)}$ for simplicity, and for $g^{(1)}$ terms, we estimate $g_{01,10}^{(0)}$ to be $\pi/t_{\rm gate}$ as mentioned before, then multiply by a factor of 1 or $\sqrt{2}$ for estimation of other terms based on the relations given in \eqref{eq:coupling}. As the plot shows, although the theoretical estimation does not give a exact agreement, it still demonstrates the overall tendency of $ZZ$ phase curve.

\section{Local equivalence class for $\sqrt{i\mathrm{SWAP}}$ gate induced by one-tone pulses}\label{App:local_rot}

In the reduced subspace of $\ket{01}$ and $\ket{10}$, the off-resonant Rabi oscillation results in a unitary evolution of the form
\begin{widetext}
\begin{equation}
    \hat{U}_{\mathrm{Rabi,\ off}}=\begin{pmatrix}
    e^{i\delta t_{\rm gate}/2}\left[\cos\left(\dfrac{\Omega t_{\rm gate}}{2}\right)-i\dfrac{\delta}{\Omega}\sin\left(\dfrac{\Omega t_{\rm gate}}{2}\right)\right] & ie^{i\delta t_{\rm gate}/2}\dfrac{g_{\rm eff}}{\Omega}\sin\left(\dfrac{\Omega t_{\rm gate}}{2}\right)\\
    ie^{-i\delta t_{\rm gate}/2}\dfrac{g_{\rm eff}}{\Omega}\sin\left(\dfrac{\Omega t_{\rm gate}}{2}\right) & e^{-i\delta t_{\rm gate}/2}\left[\cos\left(\dfrac{\Omega t_{\rm gate}}{2}\right)+i\dfrac{\delta}{\Omega}\sin\left(\dfrac{\Omega t_{\rm gate}}{2}\right)\right]
    \end{pmatrix}\,,
\end{equation}
\end{widetext}
where $\Omega=\sqrt{g_{\rm eff}^2+\delta^2}$. Notice that the diagonal terms have exactly equal and opposite phases
\begin{widetext}
\begin{equation}
    \theta_{01}=-\theta_{10}=\dfrac{\delta t_{\rm gate}}{2}-\arg\left[\cos\left(\dfrac{\Omega t_{\rm gate}}{2}\right)-i\dfrac{\delta}{\Omega}\sin\left(\dfrac{\Omega t_{\rm gate}}{2}\right)\right]\,,
\end{equation}    
\end{widetext}
hence performing single qubit $Z$-rotation on both qubits by equal and opposite angles $\theta_{01}$ and $\theta_{10}$ will bring the diagonal terms to have zero phases, without affecting $\ket{00}$ or $\ket{11}$ state. Also notice that such single qubit rotations remove the phase term $\delta t_{\rm gate}/2$, thus the phase factor $e^{i\delta t_{\rm gate}/2}$ will disappear in the local equivalent form. After this step, assuming that the only prominent process other than the off-resonant Rabi oscillation is the $ZZ$-rotation term, then the unitary evolution is equivalent to the unitary operator below when \eqref{eq:swappingcondition} is satisfied:
\begin{equation}
    \hat{U}_{\sqrt{i\mathrm{SWAP}}}=\begin{pmatrix}
    e^{-i\zeta/2} & 0 & 0 & 0\\
    0 & \dfrac{1}{\sqrt{2}} & i\dfrac{1}{\sqrt{2}}e^{i\gamma} & 0\\
    0 & i\dfrac{1}{\sqrt{2}}e^{-i\gamma} & \dfrac{1}{\sqrt{2}} & 0\\
    0 & 0 & 0 & e^{-i\zeta/2}
    \end{pmatrix}\,,
\end{equation}
where
\begin{equation}
    \gamma=\arg\left[\cos\left(\dfrac{\Omega t_{\rm gate}}{2}\right)-i\dfrac{\delta}{\Omega}\sin\left(\dfrac{\Omega t_{\rm gate}}{2}\right)\right]\,.
\end{equation}
The operator $\hat{U}_{\sqrt{i\mathrm{SWAP}}}$ can be locally transformed into $\hat{U}_{\rm ideal}(\theta,\zeta)$ by virtual $Z$ rotation before and after the Rabi oscillation
\begin{equation}
    \hat{U}(\dfrac{\pi}{2}, \zeta)=e^{i(\hat{Z}_1-\hat{Z}_2)\gamma/4}\hat{U}_{\sqrt{i\mathrm{SWAP}}}e^{-i(\hat{Z}_1-\hat{Z}_2)\gamma/4}\,,
\end{equation}
so the phase factor $\gamma$ also disappears in the final result.

\section{Branches and Extrema of Rabi Oscillation for one-tone $\sqrt{i\mathrm{SWAP}}$ gate}\label{App:branch}

\begin{center}
\begin{figure}
    \centering
    \includegraphics[width=\textwidth]{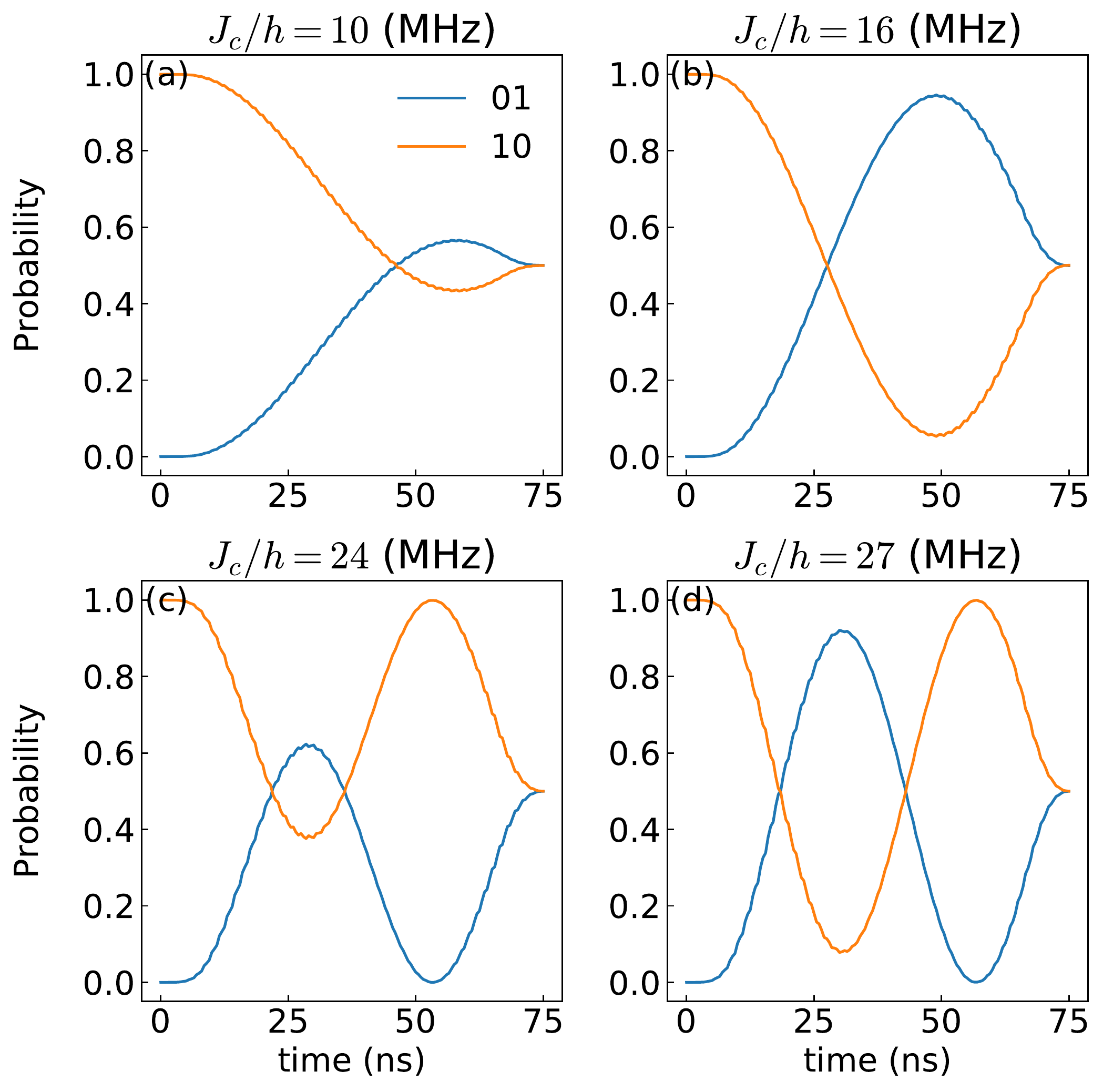}
    \caption{The probability evolution with initial state $\ket{10}$ for (a) $J_c=\SI{10}{\mega\hertz}$, (b) $\SI{16}{\mega\hertz}$, (c) \SI{24}{\mega\hertz}, (d) $\SI{27}{\mega\hertz}$. The vertical axis is the probability of measuring state $\ket{01}$ or $\ket{10}$. The result is obtained using the optimized pulse parameters of the corresponding points in Fig.~\ref{fig:sqrt_iswap_chara}.}
    \label{fig:sqrtiswapevol}
\end{figure}
\end{center}

The effectiveness of $ZZ$ phase correction is determined by the off-resonant condition \eqref{eq:swappingcondition}, and this condition has two types of extrema corresponding to the two parameters under optimization. These extrema correspond to the point in Fig.~\ref{fig:sqrtiswapevol}(e) where the plateau of zero $ZZ$ phase turns into a line of positive slope. The first type of extrema happens at $\delta=0$, as shown in Fig.~\ref{fig:sqrtiswapevol} (b) and (d), where we plot the probability of measuring the $\ket{01}$ and $\ket{10}$ state over the duration of the pulses, with the initial state to be $\ket{10}$. The Rabi oscillations for both cases are close to resonance, with the difference being that panel (b) corresponds to a $3\pi/2$ rotation, while panel (d) corresponds to a $5\pi/2$ rotation. This suggests that when we make a discontinuous change in the pulse parameters during the optimization for $\sqrt{i\mathrm{SWAP}}$ gate around $J_C=\SI{20}{\mega\hertz}$, we are actually switching to the branch of Rabi oscillation with a different period.

\newpage

\begin{center}
\begin{figure}
    \centering
    \includegraphics[width=\textwidth]{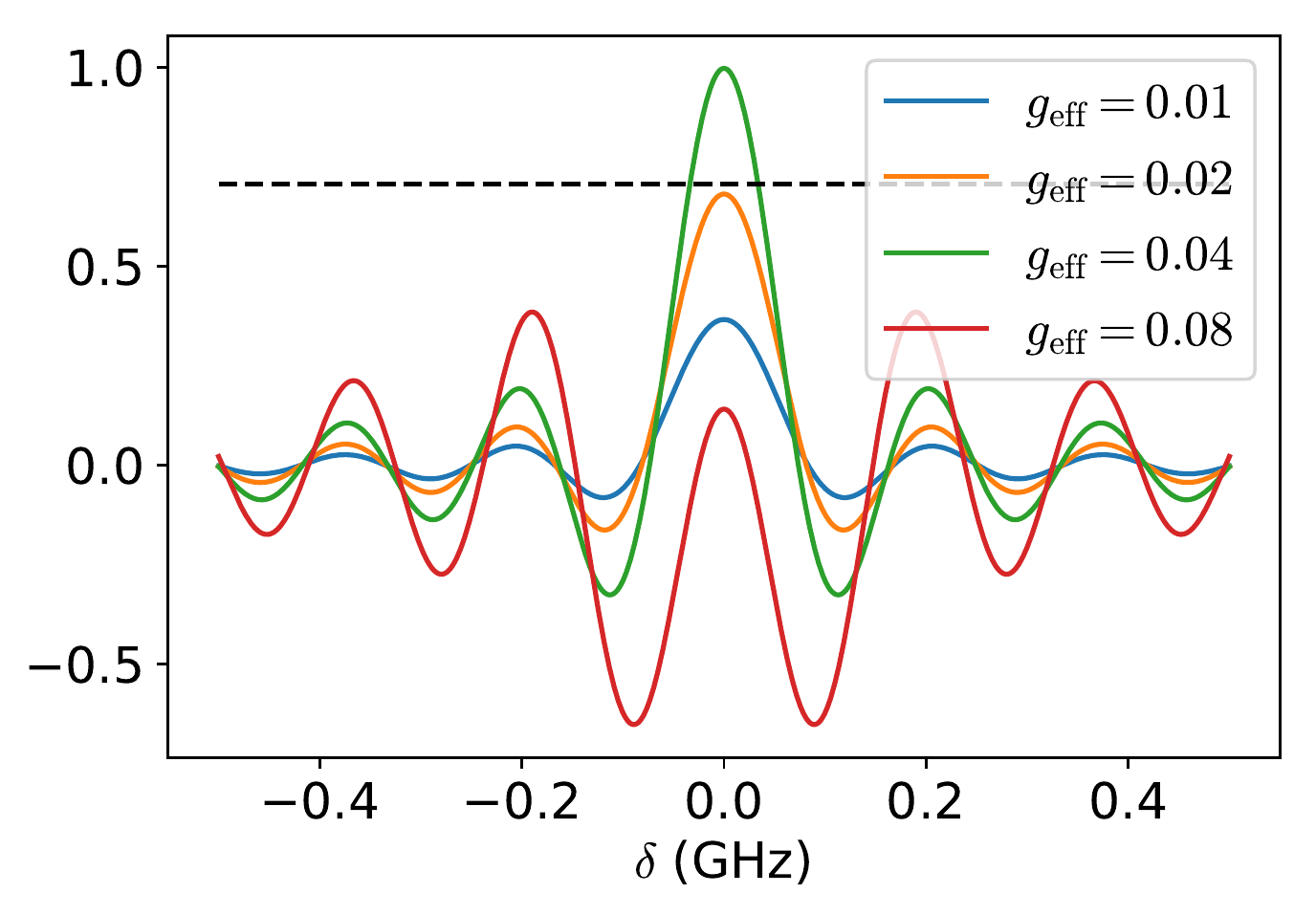}
    \caption{Evaluation of the term on the left-hand side of the condition \eqref{eq:swappingcondition} for different $g_{01,10}$. The black dashed line indicates where the curves cross $1/\sqrt{2}$ (the $-1/\sqrt{
    2}$ line is out of the range of the plot). For the detuning range $\abs{\delta}\leq\SI{0.5}{\giga\hertz}$, the curve does not necessarily reach to the value $\pm1/\sqrt{2}\approx0.707$.}
    \label{fig:rabiosci}
\end{figure}
\end{center}

The second type of extrema is related to the parameter $g_{\rm eff}$, as shown in Fig.~\ref{fig:rabiosci}. While specific values of $g_{01,10}$ might help correct the $ZZ$ phase error, the maximum swapping probability for this particular $g_{01,10}$ can be less than 1/2, making the swapping insufficient. The optimization then will prioritize condition \eqref{eq:swappingcondition}, leaving $ZZ$ phase error to be only partially corrected. This case is shown in panel (c) of Fig.~\ref{fig:sqrtiswapevol} where the amplitude of the probability oscillation is close to 1/2.

\bibliography{reference}

\end{document}